\documentclass[
reprint,
superscriptaddress,
preprintnumbers,
bibnotes,
amsmath,
amssymb,
aps,
prb,
floatfix
]{revtex4-2}

\usepackage{graphicx}
\usepackage{caption}
\usepackage{subcaption}
\usepackage{array}
\usepackage{siunitx}
\usepackage{dcolumn}
\usepackage{bm}
\usepackage{hyperref}
\hypersetup{colorlinks=true,
citecolor=blue,
linkcolor=purple,
anchorcolor=black,
urlcolor=purple}
\usepackage{mathrsfs}
\usepackage[T1]{fontenc}
\usepackage{color} 
\usepackage{braket} 
\usepackage{changes}
\definecolor{red}{rgb}{0,0,0}

\newcommand{\CoFirstAuthor}{\thanks{These authors contributed equally to this work.}}

\AtBeginDocument{
  \setlength\abovedisplayskip{10pt}
  \setlength\belowdisplayskip{10pt}
}
\begin{document}
\title{On-demand shaped photon emission based on a parametrically modulated qubit}

\author{Xiang Li}
\CoFirstAuthor
\affiliation{Beijing National Laboratory for Condensed Matter Physics, Institute of Physics, Chinese Academy of Sciences, Beijing 100190, China}
\affiliation{School of Physical Sciences, University of Chinese Academy of Sciences, Beijing 100049, China}

\author{Sheng-Yong Li}
\CoFirstAuthor
\affiliation{Department of Automation, Tsinghua University, Beijing 100084, China}

\author{Si-Lu Zhao}
\CoFirstAuthor
\affiliation{Beijing National Laboratory for Condensed Matter Physics, Institute of Physics, Chinese Academy of Sciences, Beijing 100190, China}
\affiliation{School of Physical Sciences, University of Chinese Academy of Sciences, Beijing 100049, China}

\author{Zheng-Yang Mei}
\affiliation{Beijing National Laboratory for Condensed Matter Physics, Institute of Physics, Chinese Academy of Sciences, Beijing 100190, China}
\affiliation{School of Physical Sciences, University of Chinese Academy of Sciences, Beijing 100049, China}

\author{Yang He}
\affiliation{Beijing National Laboratory for Condensed Matter Physics, Institute of Physics, Chinese Academy of Sciences, Beijing 100190, China}
\affiliation{School of Physical Sciences, University of Chinese Academy of Sciences, Beijing 100049, China}

\author{Cheng-Lin Deng}
\affiliation{Beijing National Laboratory for Condensed Matter Physics, Institute of Physics, Chinese Academy of Sciences, Beijing 100190, China}
\affiliation{School of Physical Sciences, University of Chinese Academy of Sciences, Beijing 100049, China}

\author{Yu Liu}
\affiliation{Beijing National Laboratory for Condensed Matter Physics, Institute of Physics, Chinese Academy of Sciences, Beijing 100190, China}
\affiliation{School of Physical Sciences, University of Chinese Academy of Sciences, Beijing 100049, China}

\author{Yan-Jun Liu}
\affiliation{Beijing National Laboratory for Condensed Matter Physics, Institute of Physics, Chinese Academy of Sciences, Beijing 100190, China}
\affiliation{School of Physical Sciences, University of Chinese Academy of Sciences, Beijing 100049, China}

\author{Gui-Han Liang}
\affiliation{Beijing National Laboratory for Condensed Matter Physics, Institute of Physics, Chinese Academy of Sciences, Beijing 100190, China}
\affiliation{School of Physical Sciences, University of Chinese Academy of Sciences, Beijing 100049, China}

\author{Jin-Zhe Wang}
\affiliation{Beijing National Laboratory for Condensed Matter Physics, Institute of Physics, Chinese Academy of Sciences, Beijing 100190, China}
\affiliation{School of Physical Sciences, University of Chinese Academy of Sciences, Beijing 100049, China}

\author{Xiao-Hui Song}
\affiliation{Beijing National Laboratory for Condensed Matter Physics, Institute of Physics, Chinese Academy of Sciences, Beijing 100190, China}
\affiliation{Hefei National Laboratory, Hefei 230088, China}

\author{Kai Xu}
\affiliation{Beijing National Laboratory for Condensed Matter Physics, Institute of Physics, Chinese Academy of Sciences, Beijing 100190, China}
\affiliation{Hefei National Laboratory, Hefei 230088, China}
\affiliation{Beijing Academy of Quantum Information Sciences, Beijing 100193, China}
\affiliation{CAS center for Excellence in Topological Quantum Computation, University of Chinese Academy of Sciences, Beijing 100190, China}

\author{\textcolor{red}{Heng Fan}}
\email{hfan@iphy.ac.cn}
\affiliation{Beijing National Laboratory for Condensed Matter Physics, Institute of Physics, Chinese Academy of Sciences, Beijing 100190, China}
\affiliation{School of Physical Sciences, University of Chinese Academy of Sciences, Beijing 100049, China}
\affiliation{Hefei National Laboratory, Hefei 230088, China}
\affiliation{Beijing Academy of Quantum Information Sciences, Beijing 100193, China}
\affiliation{CAS center for Excellence in Topological Quantum Computation, University of Chinese Academy of Sciences, Beijing 100190, China}

\author{Yu-Xiang Zhang}
\email{iyxz@iphy.ac.cn}
\affiliation{Beijing National Laboratory for Condensed Matter Physics, Institute of Physics, Chinese Academy of Sciences, Beijing 100190, China}
\affiliation{Hefei National Laboratory, Hefei 230088, China}

\author{Zhong-Cheng Xiang}
\email{zcxiang@iphy.ac.cn}
\affiliation{Beijing National Laboratory for Condensed Matter Physics, Institute of Physics, Chinese Academy of Sciences, Beijing 100190, China}
\affiliation{Hefei National Laboratory, Hefei 230088, China}
\affiliation{CAS center for Excellence in Topological Quantum Computation, University of Chinese Academy of Sciences, Beijing 100190, China}

\author{Dong-Ning Zheng}
\email{dzheng@iphy.ac.cn}
\affiliation{Beijing National Laboratory for Condensed Matter Physics, Institute of Physics, Chinese Academy of Sciences, Beijing 100190, China}
\affiliation{School of Physical Sciences, University of Chinese Academy of Sciences, Beijing 100049, China}
\affiliation{Hefei National Laboratory, Hefei 230088, China}
\affiliation{CAS center for Excellence in Topological Quantum Computation, University of Chinese Academy of Sciences, Beijing 100190, China}
\date{\today}

\begin{abstract}
In the circuit quantum electrodynamics architectures, to realize a long-range quantum network mediated by flying photon, it is necessary to shape the temporal profile of emitted photons to achieve a high transfer efficiency between two quantum nodes. In this work, we demonstrate a new single-rail and dual-rail time-bin shaped photon generator without additional flux-tunable elements, which can act as a quantum interface of a point-to-point quantum network. In our approach, we adopt a qubit-resonator-transmission line configuration, and the effective coupling strength between the qubit and the resonator can be varied by parametrically modulating the qubit frequency. In this way, the coupling is directly proportional to the parametric modulation amplitude and covers a broad tunable range beyond \SI{20}{MHz} for the sample we used. Additionally, when emitting shaped photons, we find that the spurious frequency shift (\SI{-0.4}{MHz}) due to parametric modulation is small and can be readily calibrated through chirping. We develop an efficient photon field measurement setup based on the data stream processing of GPU. Utilizing this system, we perform photon temporal profile measurement, quantum state tomography of photon field and quantum process tomography of single-rail quantum state transfer based on heterodyne measurement scheme. The single-rail encoding state transfer fidelity of shaped photon emission is $90.32\%$, and that for unshaped photon is $97.20\%$, respectively. We believe that the fidelity of shaped photon emission is mainly limited by the qubit coherence time. The results demonstrate that our method is hardware efficient, simple to implement and scalable. It could become a viable tool in a high-quality quantum network utilizing both single-rail and dual-rail time-bin encoding.
\end{abstract}
\maketitle

\section{Introduction}
\label{sec:1}
    In the realm of quantum networking, it is necessary to coherently link various quantum nodes such as quantum processors, memories, and sensors together~\cite{kimble2008quantum,wehner2018quantum,awschalom2021development}. The core of a quantum network is the transmission of quantum states and the generation of remote quantum entanglement. In optical contexts, due to the low efficiency of single-photon generation, one typically generates heralded entanglement probabilistically via coincidence counts of flying photons from independent sources~\cite{yurke1992bell,barrett2005efficient,hensen2015loophole}. In contrast, in the domain of superconducting quantum computing, circuit QED architectures~\cite{blais2007quantum,blais2020quantum,blais2021circuit} have enabled substantial coupling efficiency between superconducting qubits, microwave resonators, and 1D waveguides. The microwave photon field emitted by the qubits is confined in 1D waveguides, eliminating the need for spatial mode matching, which is key to on-demand, high-efficiency quantum state transfer. Many recent experiments are based on Cirac-Zoller-Kimble-Mabuchi (CZKM) schemes~\cite{cirac1997quantum}. Therein, single photons are pitched and caught directly by symmetrically shaping and time-reversal quantum control in emission and receiving nodes~\cite{korotkov2011flying,sete2015robust,yao2005theory,vogell2017deterministic}.

    Regarding the devices used in microwave photon emission and receiving, there are two primary configurations: In the first case, one directly couples quantum emitters to 1D waveguides~\cite{yin2013catch,wenner2014catching,forn2017demand,zhong2019violating,chang2020remote,burkhart2021error,zhong2021deterministic,yan2022entanglement,niu2023low,qiu2023deterministic,grebel2023bidirectional}, while in the other case one employs a buffer resonator or qubit for indirect coupling~\cite{pechal2014microwave,srinivasan2014time,kurpiers2018deterministic,kurpiers2019quantum,magnard2020microwave,storz2023loophole,pfaff2017controlled,axline2018demand,yang2023deterministic,besse2020realizing}. The former can in principle, release and catch shaped photons more rapidly and is exposed to fewer dissipation channels. However, most implementations~\cite{yin2013catch,wenner2014catching,zhong2019violating,chang2020remote,zhong2021deterministic,yan2022entanglement,niu2023low,qiu2023deterministic,grebel2023bidirectional} require a flux-tunable loop of Josephson junctions to adjust the mutual inductance between the emitter and the waveguide, causing significant frequency shifts in the emitter that are challenging to calibrate accurately. To compare, with the aid of the extra resonator (qubit), the configurations with buffer support indirect tunable coupling like cavity-assisted Raman process~\cite{zeytinouglu2015microwave} and dual-rail encoding protocols that are more robust against photon transfer loss~\cite{kurpiers2019quantum,li2023frequency} and phase reference error between sender and receiver~\cite{ilves2020demand}. The drawback is therefore the additional loss induced by the buffer and the limited bandwidth of the emitted photon which is constrained by both the coupling strength between the quantum emitters and the buffer and the decay rate of the buffer~\cite{magnard2021meter}. For all the methods mentioned above, quantum emitters experience unintended frequency shifts while photon shaping. Additionally, the relationship between effective coupling strength and control pulse amplitude is either non-linear~\cite{yin2013catch,forn2017demand,zhong2019violating} or has a constrained linear range~\cite{pechal2014microwave,kurpiers2018deterministic,kurpiers2019quantum}.

   In this work, we propose and experimentally demonstrate a new shaped photon generator employing a buffer emission resonator. We modulate the qubit parametrically by applying a time-varying magnetic flux and thus use the first-order side-band of a parametrically modulated qubit to induce an effective coupling between the transmon qubit and the emission resonator. The coupling strength is found to be proportional to the parametric modulation pulse amplitudes. The spurious frequency shift of the qubit is small, attributed to the modest amplitude of the necessary parametric modulation pulse. Consequently, the photon shaping and frequency shift calibration processes are straightforward and easy to implement experimentally. It avoids the use of any additional flux-tunable components with Josephson junctions and serves as a quantum interface functioning as both the sender and receiver in a point-to-point quantum network. Additionally, with the resonator, it is capable of single-rail and dual-rail time-bin photon emission. Through quantum process tomography, we achieve a fidelity of single-rail quantum state transfer with shaped photon at 90.32\% and with unshaped photon at 97.20\%.

\section{flux-modulation of a frequency tunable qubit}
\label{sec:2}
    \begin{figure}[htbp]
        \centering
        \includegraphics[width=0.45\textwidth]{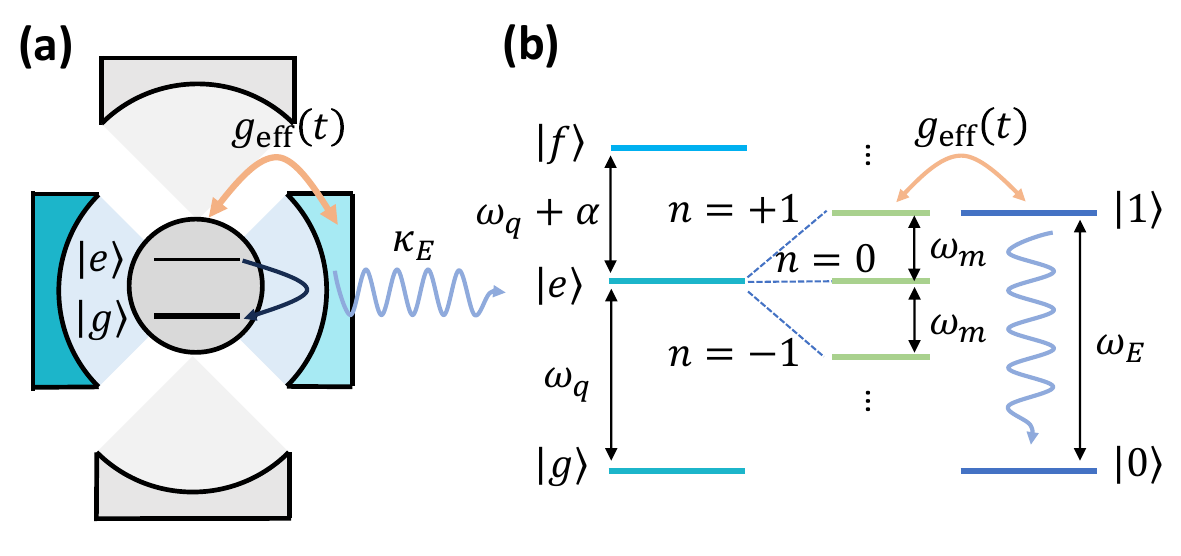}
        \caption{\textbf{Schematic diagrams of sample layouts and energy levels.} \textbf{(a)}Our sample layout which has two resonators coupled to the qubit, one for dispersive readout, the other for shaped photon emission. \textcolor{red}{The effective coupling $g_{\text{eff}}(t)$ between the qubit and emission resonator is controlled by parametric modulation. Decay rate of the emission resonator is denoted by $\kappa_E$}. \textbf{(b)}The energy levels of transmon and emission resonator \textcolor{red}{where $\omega_q$ is the angular frequency of transmon, $\alpha$ is the anharmonicity of transmon and $\omega_E$ is the angular frequency of emission resonator}. \textcolor{red}{Modes of readout resonator are not shown explicitly. The light blue levels represent the lowest three level of the transmon, the green levels represent the zeroth and first order sidebands of parametric modulation, and the dark blue levels represent the zero photon and one photon levels of the emission resonator.} Using the first order side-band of parametric modulation we couple the transmon qubit to emission resonator and modulate the amplitude dynamically to shape the emitted photon.}
        \label{fig:1}
    \end{figure}

    Our scheme includes a flux modulated frequency tunable superconducting transmon qubit capacitively coupled to two coplanar waveguide resonators, one for dispersive readout, the other one for shaped photon emission, as schematically shown in Fig.~\ref{fig:1}(a).  We use a sinusoidal parametric modulation pulse added to the DC-SQUID loop of transmon $\Phi(t)=\Phi_{\text{DC}}+\Phi_{\text{AC}}\cos(\omega_{m}t+\theta_{m})$, where $\Phi_{\text{DC}}$ is the static flux bias, $\omega_{m}$ is the frequency of parametric modulation and $\theta_{m}$ is the initial phase of parametric modulation. For a frequency tunable transmon, we have $\omega_q(\Phi) = \frac{1}{\hbar} \left(\sqrt{8E_{\text{C}} E_{\text{J}}(\Phi)} - E_{\text{C}}\right)$ where $E_{\text{J}}(\Phi) = E_{\text{J}\Sigma} \cos\left(\pi \Phi/\Phi_0\right)$, $E_{\text{J}\Sigma}$ is the total Josephson energy, \textcolor{red}{$E_{\text{C}}$ is the capacitive energy of the transmon, $\Phi_{0}$ denotes the magnetic flux quantum.} Due to the non-linear relationship between the magnetic flux and transmon frequency, the parametric modulation leads to the variation of qubit frequency consists of different frequency components $\omega_{q}(t)=\bar{\omega}_{q}+ \sum_{a=1}^{\infty} A^{a}_{m}\cos[a(\omega_{m}t+\theta_{m})](a \in \mathbb{N})$~\cite{didier2018analytical,zhou2021rapid}, where $\bar{\omega}_{q}$ is the average frequency of qubit during parametric modulation, $A^{a}_{m}$ is the $ath$ order Fourier components of longitudinal field modulation (LFM) amplitude (See Appendix.~\ref{sec:app3}). When a weak parametric modulation pulse ($A^{1}_{m} \ll \omega_{m}$) which has a frequency of $\omega_{m} \approx \omega_{E}-\omega_{q}$ is applied to the qubit, \textcolor{red}{where $\omega_{E}$ is the frequency of the emission resonator, $\omega_{q}$ is the frequency of qubit,} neglecting the far detuning terms, the interaction Hamiltonian under two-level approximation is
    \begin{align}
    \label{eq:1}
    \hat{H}_{\textcolor{red}{\text{int}}}/\hbar&=g_{qE}J_{1}(\frac{A^{1}_{m}}{\omega_{m}})e^{i\theta_m}\hat{a}\hat{\sigma}_{+} + h.c. \nonumber \\
    &\approx g_{qE}\frac{A^{1}_{m}}{2\omega_{m}}e^{i\theta_m}\hat{a}\hat{\sigma}_{+} + h.c. \nonumber \\
    &\approx \frac{g_{qE}}{2\omega_m} \Phi_{\text{AC}}\left.\frac{d\omega_q}{d\Phi}\right|_{\Phi_{\text{DC}}}e^{i\theta_{m}}\hat{a}\hat{\sigma}_{+} + h.c.,
    \end{align}
     where \textcolor{red}{$J_1(x)$ is the first order Bessel function (when $x \to 0$, $J_1(x) \approx x/2$), } $g_{qE}$ is the coupling coefficient between the qubit and emission resonator, $A^{1}_{m}$ is the first order Fourier components of LFM amplitudes, $\theta_{m}$, here, becomes the phase angle of the first-order side-band coupling. Equation(\ref{eq:1}) demonstrates an effective first-order side-band coupling between the qubit and emission resonator $g_{\textcolor{red}{\text{eff}}}=g_{qE}A^{1}_{m}/(2\omega_{m})e^{i\theta_{m}} \propto \Phi_{\text{AC}} e^{i\theta_{m}}$ that is linear to the parametric modulation pulse amplitude $\Phi_{\text{AC}}$ and the phase can be controlled by the modulation pulse $\theta_{m}$. Notice that the coupling is the first-order coupling and as a result we can get a wide range of tunable coupling with a weak parametric modulation pulse.
     When we apply a weak parametric modulation with a time-dependent amplitude and phase $\Phi(t)=\Phi_{\text{DC}}+\Phi_{\text{AC}}(t)\cos(\omega_{m}t+\theta_{m}(t))$, the first order effective coupling will also gain time-dependent amplitude and phase which is linear to the parametric modulation pulse. 
    \begin{align}
    g_{\textcolor{red}{\text{eff}}}(t)&= \frac{g_{qE}}{2\omega_m} \Phi_{\text{AC}}(t)\left.\frac{d\omega_q}{d\Phi}\right|_{\Phi_{\text{DC}}}e^{i\theta_{m}(t)} \nonumber \\
    &\propto \Phi_{\text{AC}}(t) e^{i\theta_{m}(t)}.
    \end{align}
    Thus, we can get arbitrary emitted photon shape~\cite{Gorshkov2007Photon} through control the envelope and phase of parametric modulation pulses.

    In addition, it should be noted that parametric modulation itself results in qubit frequency shift $\Delta = \Delta_{\text{DC}} + \Delta'_{\text{Lamb}}$, where $\Delta_{\text{DC}}=\overline{\omega}_{q}-\omega_{q}(\Phi_{\text{DC}})$ is the average frequency shift during parametric modulation and $\Delta'_{\text{Lamb}}$ is the Lamb shift variation induced by the zero-order side-band~\cite{naik2017random} (See Appendix.~\ref{sec:app3}). The frequency shift is approximately quadratic to parametric modulation amplitude. Therefore, we can experimentally characterize(See Appendix.~\ref{sec:app4}) and calibrate the frequency shift with a chirp pulse $\Phi_{m}(t)=\Phi_{\text{AC}}(t)\cos(\omega_{m}t+\theta_{m}(t))$ where $\dot{\theta}_{m}(t)=-\Delta(t)$.
    
    For our device, the peak frequency of transmon is $\omega_{q,p}/2\pi=5.997$ GHz, and the frequency of readout and emission resonator is $\omega_{R,p}/2\pi=4.177$ GHz and $\omega_{E,p}/2\pi=7.142$ GHz, respectively. During the experiment, we bias our transmon qubit slightly away from the peak frequency using a static external magnetic flux, setting \(\Phi_{\text{DC}}/\Phi_{0} = 0.04\) where \(\Phi_{0}\) denotes the magnetic flux quantum. By doing so, the qubit maintains a relatively high coherence time \(T^{*}_{2}\), meanwhile, we can  circumvent the undesired simultaneous coupling of the first and the second side-bands. At that point, the frequency of qubit and emission resonator is $\omega_{q}/2\pi=5.947$ GHz and $\omega_{E}/2\pi=7.139$ GHz, respectively. Emission resonator frequency shifts due to the Lamb shift induced by large dispersive coupling between the qubit and emission resonator. We then perform weak parametric modulation pulses to use the effective coupling of the first-order side-band of the qubit and emission resonator to shape the photon temporal profile and characterize the photons.

\section{Photon shaping and dynamical frequency calibration by chirping}
\label{sec:3}
\begin{figure*}[htbp]
        \centering
        \includegraphics[width=0.9\textwidth]{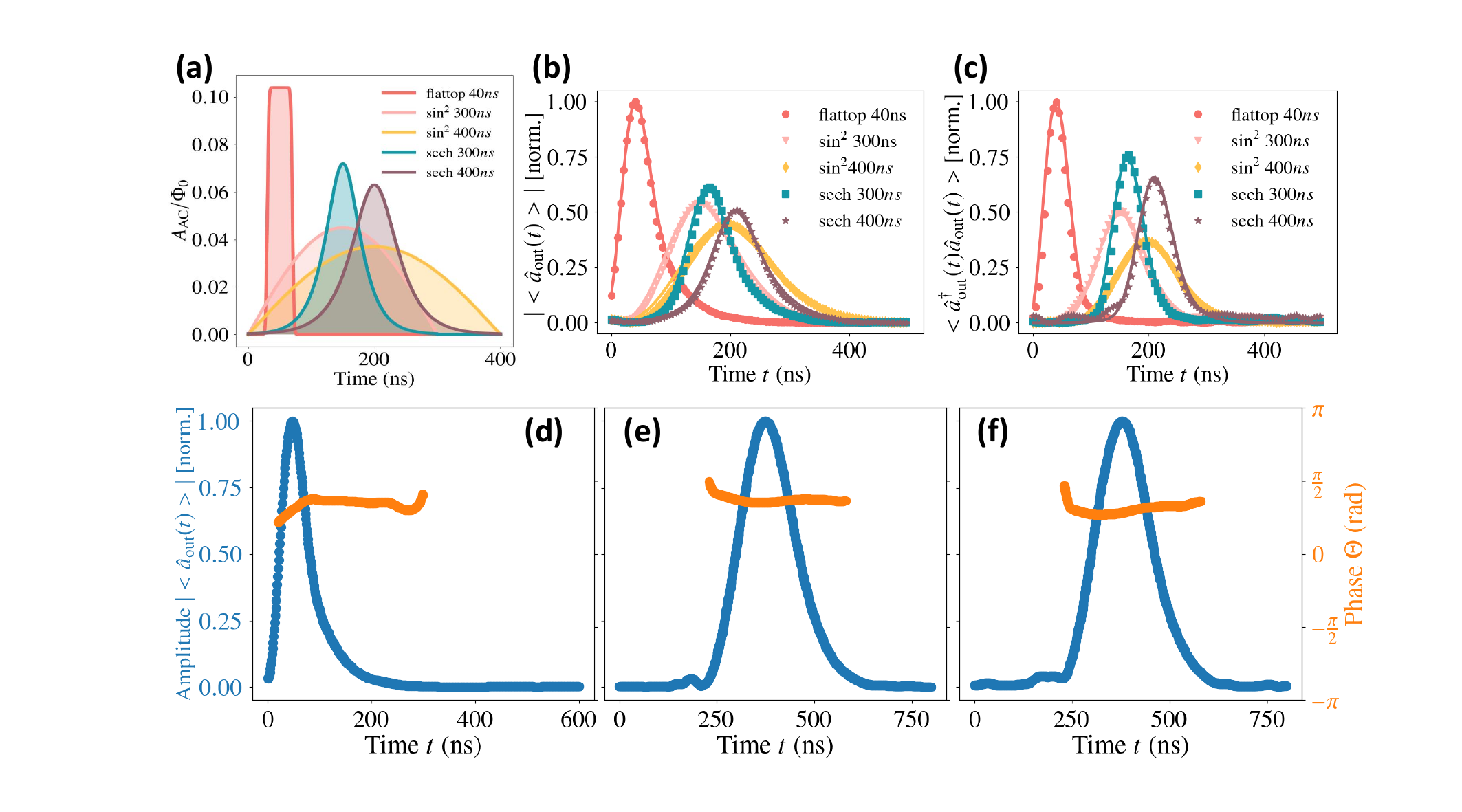}
        \caption{\textbf{Photon temporal profile of $(\ket{0}+\ket{1})/\sqrt{2}$ state using different control pulses} (flattop, $\sin^{2}$, and sech envelopes). Flattop pulse generates the unshaped photon without frequency calibration by chirping, other pulses generate the shaped photons with frequency calibration by chirping. \textcolor{red}{The measured data is averaged for $3 \times 10^6$ times.} \textcolor{red}{\textbf{(a)}Different parametric modulation pulses ($\pi_{e0g1}$ pulses) used to eliminate the excitation of qubit and generate photon.} \textbf{(\textcolor{red}{b})}The amplitude of the output photon field. The dot is the experimental data, the solid lines are the numerical simulations using QuTiP~\cite{JOHANSSON20121760}. \textbf{(\textcolor{red}{c})}The power of output photon field. The dot is the experimental data, the solid lines are the numerical simulations using QuTiP. \textbf{(\textcolor{red}{d})}The amplitude and phase of unshaped photon generated by \SI{40}{\nano\second} flattop pulse. \textbf{(\textcolor{red}{e})}The amplitude and phase of shaped photon generated by \SI{400}{\nano\second} $\sin^{2}$ pulse with frequency calibration by chirping. \textbf{(\textcolor{red}{f})}The amplitude and phase of shaped photon generated by \SI{400}{\nano\second} $\sin^{2}$ pulse without frequency calibration by chirping.}
        \label{fig:2}
    \end{figure*}

     \textcolor{red}{For shaped photon generation, we excite the qubit by conventional charge drive, control the effective coupling between the qubit and emission resonator. Then, shaped photon is leaked out from the emission resonator. Due to number-phase uncertainty, the temporal envelop cannot be determined experimental from photon field in the Fock state such as $\ket{1}$. To observe the envelope of the photon, we need to prepare a state with uncertain photon numbers, such as $(\ket{0}+\ket{1})/\sqrt{2}$. So,} we first excite the qubit with a ${\pi/2}_{ge}$ pulse from $\ket{g}$ to $(\ket{g}+\ket{e})/\sqrt{2}$ then use the first-order side-band parametric modulation pulse to eliminate the qubit $\ket{e}$ state population \textcolor{red}{and emit a half photon state $(\ket{0}+\ket{1})/\sqrt{2}$ from the emission resonator} and obtain a so-called $\pi_{e0g1}$ pulse. Because our qubit has a negligible thermal excitation (about 0.2\% $\ket{e}$ population, see Appendix.~\ref{sec:app7}), we perform the experiment without extra qubit initialization. Then, using heterodyne detection method~\cite{Silva2010schemes} with a linear amplification chain (with a Josephson parametric amplifier (JPA)~\cite{yamamoto2008flux,roy2015broadband} and the quantum efficiency is $\eta=0.26$, see Appendix.~\ref{sec:app5}) and a linear voltage detector (a high speed data acquisition card with the data stream processing of GPU) (See Appendix.~\ref{sec:app2}) we derive the time-dependent quadrature voltages and power of the photon field. We use different pulse envelopes of parametric modulation pulses \textcolor{red}{(See Fig.~\ref{fig:2}(a))} such as flattop pulse which is square pulse convoluted with Gaussian function with $2\sigma=4ns$ to have a smooth rising edge, $\sin^{2}$ pulse which has a form of $\Phi_{\text{AC}}(t)=A_{\text{AC}}\sin(\pi t/T)$ and no need to truncate and sech pulse which has the form of $\Phi_{\text{AC}}(t)=A_{\text{AC}}\text{sech}(\kappa t)$ truncated to the length of $4\pi/\kappa$. As a comparison, flattop pulse generates the unshaped photon without frequency calibration by chirping, while other pulses generate the shaped photons with frequency calibration by chirping. The average amplitude results are shown in Fig.~\ref{fig:2}(\textcolor{red}{b}), the average power results are shown in Fig.~\ref{fig:2}(\textcolor{red}{c}).
    
    Following Ref.~\cite{pechal2014microwave}, we apply a symmetry factor to characterize the symmetry of both amplitude and phase of photon temporal profile by \textcolor{red}{auto-correlation} of the measured quadratures normalized by the quadrature amplitudes:
    
    \begin{align}
    \label{eq:3}
    s = \displaystyle \max_{t_0} \frac{\int \braket{\hat{a}_{\textcolor{red}{\text{out}}}(t_{0}-t)}^{*} \braket{\hat{a}_{\textcolor{red}{\text{out}}}(t)} \,dt}{\int |\braket{\hat{a}_{\textcolor{red}{\text{out}}}(t)}|^{2} \, dt}.
    \end{align}
    
    In our case, the decay rate of emission resonator is about $\kappa_{E}/2\pi = 5.2$ \si{MHz}. Thus, the bandwidth upper-bond of our shaped photon is $\Gamma/2\pi \leq \kappa_{E}/2\pi = 5.2$ \si{MHz}. Truncated to $4\pi/\Gamma$, the minimal symmetrical photon time length is about \SI{385}{\nano\second}~\cite{magnard2021meter}, thus the appropriate parametric modulation pulse length is \SIrange{300}{400}{\nano\second}. We evaluate the symmetry of the unshaped photon generated by flattop pulse and shaped photon generated by $\sin^{2}$ and sech pulses in Table.~\ref{tab:1}. We experimentally find that for pulses with the same length, sech-type parametric modulation pulses always generate photons that have lower symmetry factor than that of $\sin^{2}$-type pulses. \textcolor{red}{We note that if we define a symmetry factor without phase by replacing the $\braket{\hat{a}_{\text{out}}(t_{0}-t)}$, $\braket{\hat{a}_{\text{out}}(t)}$ with $|\braket{\hat{a}_{\text{out}}(t_{0}-t)}|$ and $|\braket{\hat{a}_{\text{out}}(t)}|$ in Eq.~\ref{eq:3}, this absolute symmetry factor of photon generated by \SI{400}{\nano\second} sech-type pulse is 0.9974 while that of \SI{400}{\nano\second} $\sin^{2}$-type pulse is 0.9917 (See Table.~\ref{tab:1}) which means the absolute amplitude results of sech-type pulses are more symmetrical than $\sin^{2}$-type pulses. However,} the phase of sech-type pulses drifts more (not shown in the figure). We believe it is attributable to the distortion of the parametric modulation pulse shape, which is caused by the unknown transfer function of the flux line. The sech-type pulses have sharper peaks and larger amplitudes for the same $\pi_{e0g1}$ pulse length, thus they are more distorted and result in a spurious frequency shift that are calibrated mistakenly. As a result, the most symmetrical photon is created by $\sin^{2}$ pulse of $A_{\text{AC}}=0.037\Phi_{0}$ and \SI{400}{\nano\second} length whose symmetry factor is over 0.99 with a residue $\ket{e}$ state excitation of $0.13 \pm 0.02\%$. 
    
    \begin{table}[htbp]
    \centering
    \renewcommand{\arraystretch}{1.5}
    \begin{tabular}{|@{\hskip 0.2cm}c@{\hskip 0.2cm}|@{\hskip 0.3cm}c@{\hskip 0.3cm}|@{\hskip 0.3cm}c@{\hskip 0.3cm}|@{\hskip 0.3cm}c@{\hskip 0.3cm}|@{\hskip 0.3cm}c@{\hskip 0.3cm}|}
        \hline
        \textbf{Pulse shape} & \textbf{\textcolor{red}{$L$}} & \textbf{\textcolor{red}{$A_{\text{AC}}$}} & \textbf{\textcolor{red}{s}} & \textbf{\textcolor{red}{$s_{\text{abs}}$}} \\ \hline
        flattop & 40ns & 0.104$\Phi_{0}$ & 0.8933 & \textcolor{red}{0.9567} \\ \hline
        $\sin^{2}$ & 300ns & 0.045$\Phi_{0}$ & 0.9710 & \textcolor{red}{0.9904} \\  \hline
        $\sin^{2}$ & 400ns & 0.037$\Phi_{0}$ & 0.9906 & \textcolor{red}{0.9917} \\  \hline
        sech & 300ns & 0.072$\Phi_{0}$ & 0.9112 & \textcolor{red}{0.9864} \\  \hline
        sech & 400ns & 0.063$\Phi_{0}$ & 0.9424 & \textcolor{red}{0.9974} \\  \hline
    \end{tabular}
    \caption{\textbf{The symmetry factor of different parametric modulation pulses.} \textcolor{red}{$L$ represents the length of parametric modulation pulse, $A_{\text{AC}}$ represents the amplitude of parametric modulation pulse, $s$ represents the symmetry factor defined by Eq.~\ref{eq:3}, and $s_{\text{abs}}$ represents the symmetry factor without phase which replaces $\braket{\hat{a}_{\text{out}}(t_0 - t)}$, $\braket{\hat{a}_{\text{out}}(t)}$ with $|\braket{\hat{a}_{\text{out}}(t_0 - t)}|$ and $|\braket{\hat{a}_{\text{out}}(t)}|$ in Eq.~\ref{eq:3}.} Flattop pulse generates the unshaped photon without frequency calibration by chirping, other pulses generate the shaped photons with frequency calibration by chirping.}
    \label{tab:1}
    \end{table}

    We demonstrate the amplitude ($|\braket{\hat{a}_{\textcolor{red}{\text{out}}}(t)}|$) and phase ($\Theta(t) = arg(\braket{\hat{a}_{\textcolor{red}{\text{out}}}(t)})$) of the most symmetrically shaped photon and compare the results of that with/without frequency calibration by chirping together with unshaped photon in Fig.~\ref{fig:2}(\textcolor{red}{d-f}). We note that compared to unshaped photon (in Fig.~\ref{fig:2}(\textcolor{red}{d})) the shaped photon (in Fig.~\ref{fig:2}(\textcolor{red}{e})) has a significant symmetry enhancement, the symmetry factor is enhanced from 0.8933 to 0.9906. Compared to the phase of unshaped photon, the phase of shaped photon almost stays constant, too. However, the symmetry factor of shaped photon without frequency calibration by chirping (in Fig.~\ref{fig:2}(\textcolor{red}{f})) drops because of the phase accumulation of photon induced by dynamical frequency shift caused by varying amplitude of the parametric modulation pulse. In spite of this, it is not much, the symmetry factor is still about 0.9862 (compared to 0.9906 of the result with frequency calibration by chirping) because of the small frequency shift (about \SI{-0.4}{MHz}) of the weak parametric modulation $A_{\text{AC}}=0.037\Phi_{0}$(See Appendix.~\ref{sec:app4}). 
    
\section{photon emission with different rabi angle}
\label{sec:4}
\begin{figure*}[htbp]
        \centering
        \includegraphics[width=1\textwidth]{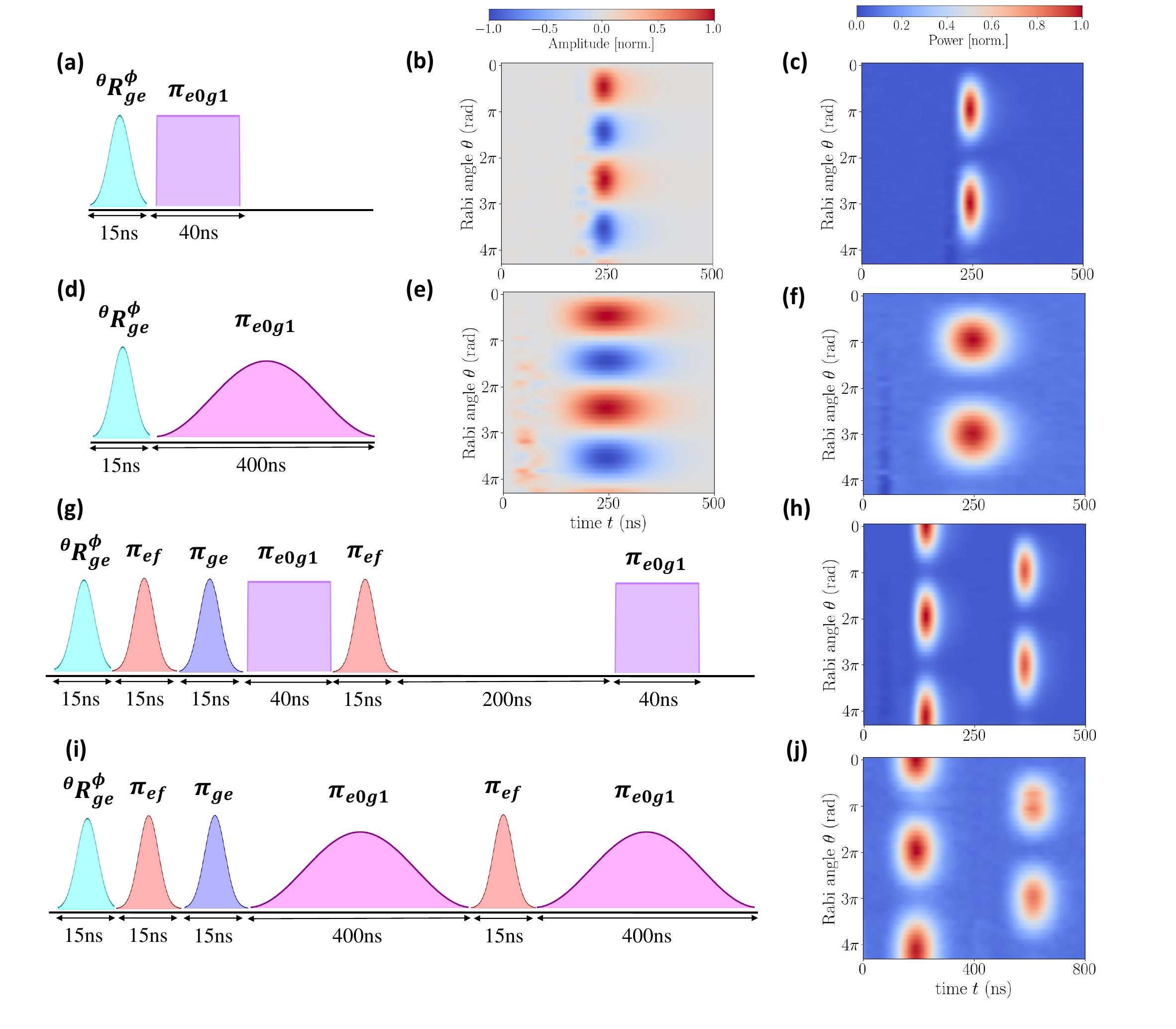}
        \caption{\textbf{Fock state photon and time-bin photon of different Rabi angles.} \textcolor{red}{The data is measured by heterodyne detection and averaged for $3 \times 10^6$ times.} \textbf{(a)}Control sequence to generate unshaped Fock state photon. \textbf{(b-c)}Amplitude ($Re\braket{\hat{a}_{\textcolor{red}{\text{out}}}(t)}$) and power ($\braket{\hat{a}^{\dagger}_{\textcolor{red}{\text{out}}}(t)\hat{a}_{\textcolor{red}{\text{out}}}(t)}$) of different Rabi angle (from 0 to $4\pi$, all the following are the same) unshaped Fock state photon. \textbf{(d)}Control sequence to generate shaped Fock state photon. \textbf{(e-f)}Amplitude ($Re\braket{\hat{a}_{\textcolor{red}{\text{out}}}(t)}$) and power ($\braket{\hat{a}^{\dagger}_{\textcolor{red}{\text{out}}}(t)\hat{a}_{\textcolor{red}{\text{out}}}(t)}$) of different Rabi angle shaped Fock state photon. \textbf{(g)}Control sequence to generate unshaped time-bin photon. \textbf{(h)}Power ($\braket{\hat{a}^{\dagger}_{\textcolor{red}{\text{out}}}(t)\hat{a}_{\textcolor{red}{\text{out}}}(t)}$) of different Rabi angle unshaped time-bin photon. \textbf{(i)}Control sequence to generate shaped time-bin photon. \textbf{(j)}Power ($\braket{\hat{a}^{\dagger}_{\textcolor{red}{\text{out}}}(t)\hat{a}_{\textcolor{red}{\text{out}}}(t)}$) of different Rabi angle shaped time-bin photon.} 
        \label{fig:3}
    \end{figure*}

    To further verify the photon emission process, we prepare our qubit in states with different Rabi angles $\theta$, $\cos{(\theta/2)}\ket{g} + \sin{(\theta/2)}e^{i\phi}\ket{e}$. In Fig.~\ref{fig:3}, we produce single-rail photon and dual-rail time-bin photon with control pulse sequences shown in Fig.~\ref{fig:3}(a),(d),(g),(i) respectively. For single-rail photon generation, we use the single qubit rotation gate $R_{ge}(\theta,\phi)$ and the parametric modulation $\pi_{e0g1}$ pulse which facilitates the process $\ket{e0} \to \ket{g1}$ to eliminate the excitation of qubit and map the quantum information to the emission resonator and finally into the propagating mode. Notably, we can identify that the amplitude $Re\braket{\hat{a}_{\textcolor{red}{\text{out}}}(t)}$ changes sign along with Rabi angle per period of $\pi$ because the off-diagonal elements of the density matrix of qubit changes sign which represent the process of photon emission maintains phase coherence. Moreover, by comparing the unshaped and shaped results in Fig.~\ref{fig:3}(b-c) and Fig.~\ref{fig:3}(e-f), the shape of the photon remains symmetric and is unaffected by variations in the Rabi angles.

    As for the dual-rail time-bin photon generation protocol(as shown in Fig.~\ref{fig:3}(g),(i)), we need $\ket{f}$ state to be an ancillary state to temporarily store the quantum information from $\ket{e}$ state for the second time-bin photon emission. Therefore, we use a $\pi_{ef}$ pulse to interchange the coefficients of qubit $\ket{e}$ state and $\ket{f}$ state, then use a $\pi_{ge}$ pulse to map $\ket{g}$ state to $\ket{e}$ state and use $\pi_{e0g1}$ pulse to map the quantum information stored in $\ket{g}$ state into the first time-bin photon. Next, we use another $\pi_{ef}$ pulse to reload the quantum information of $\ket{e}$ state from $\ket{f}$ state, and finally use the second $\pi_{e0g1}$ pulse to map quantum information stored in $\ket{e}$ state into the second time-bin photon. In general, we generate entangled states like $\cos{(\theta/2)}\ket{01} + \sin{(\theta/2)}e^{i\phi}\ket{10}$, where $\ket{01}$ represents one photon in the early time-bin mode, $\ket{10}$ represents one photon in the latter time-bin mode. In the subspace of each mode, the quantum state does not have any coherence. Thus, we neither observe non-zero quadrature amplitudes of time-bin photon nor show the amplitude results here. Results of power measurements demonstrate the quantum state mapping between the qubit and time-bin photon although the phase information is omitted. We notice that the second time-bin photon has less power than the first one because of the limited energy relaxation time of $\ket{f}$ state. Moreover, the shaped photon maintains symmetric at both the first time-bin mode and the second and is still unaffected by variations in the Rabi angles.
    
\section{Quantum state and process tomography of propagating modes}
\label{sec:5}
\begin{figure}[htbp]
        \centering
        \includegraphics[width=0.45\textwidth]{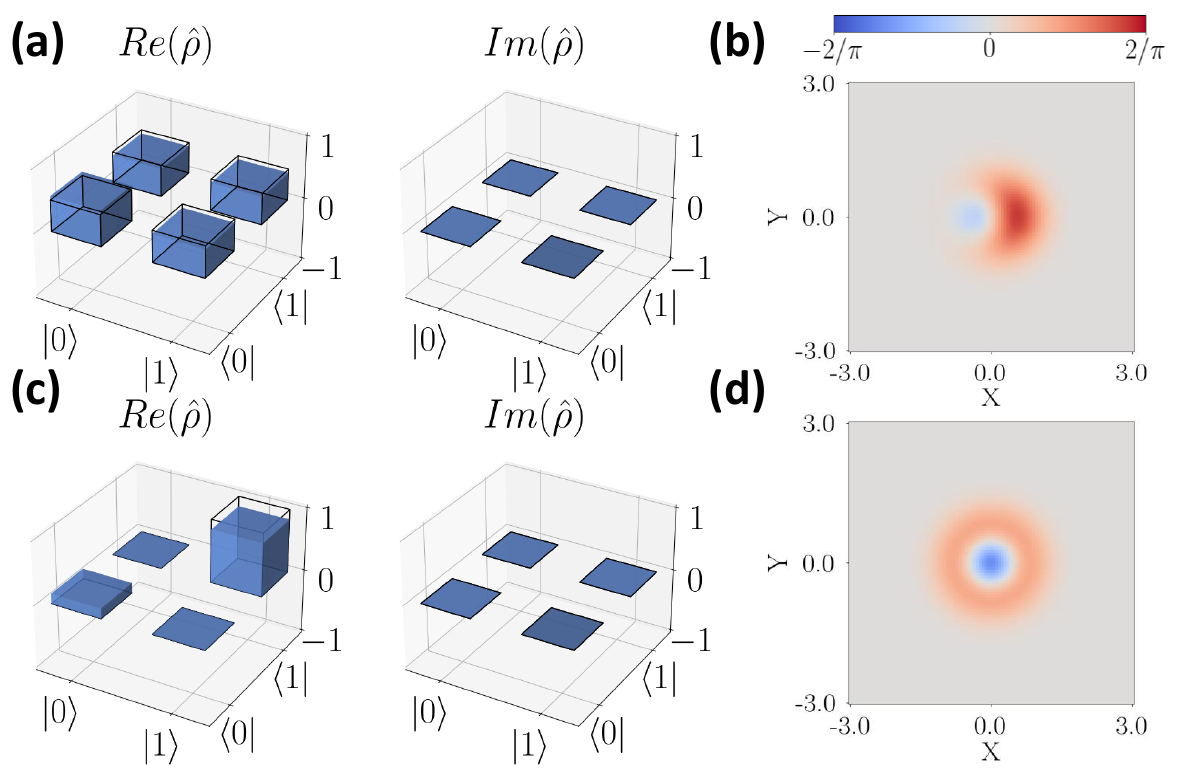}
        \caption{\textbf{Quantum state tomography of typical shaped itinerant photon state with frequency calibration by chirping.} \textcolor{red}{The measured data is averaged for $3 \times 10^7$ times.} \textbf{(a)}The density matrix of generated $(\ket{0}+\ket{1})/\sqrt{2}$ state. \textbf{(b)}Wigner function of generated $(\ket{0}+\ket{1})/\sqrt{2}$ state. \textbf{(c)}The density matrix of generated $\ket{1}$ state. \textbf{(d)}Wigner function of generated $\ket{1}$ state.}
        \label{fig:4}
    \end{figure}

    We then perform quantum state tomography~\cite{eichler2012characterizing,eichler2013experimental} to characterize the state fidelity of quantum state of \textcolor{red}{the most symmetrical shaped photon ($\text{sin}^2$ 400ns pulse)} of our protocol. We use the matched filter of the same shape of $(\ket{0}+\ket{1})/\sqrt{2}$ state quadratures mentioned above to maximize the detection efficiency. Then we generate the 'ON' histogram containing both signal and noise ($\hat{S}_{ON}=\hat{a}+\hat{h}^{\dagger}$) from the amplification chain and the 'OFF' histogram only contains noise ($\hat{S}_{OFF}=\hat{h}^{\dagger}$) from the amplification chain from $3 \times 10^{7}$ repeats of experiments. Next, we estimate the gain from the amplification chain using the relation $\braket{\hat{a}^{\dagger}\hat{a}}=|\braket{\hat{a}}|$ for $1/\sqrt{2}(\ket{0}+\ket{1})$ state. Finally, we extract the signal moments $\braket{(\hat{a}^{\dagger})^{n}\hat{a}^{m}}$ ($n,m \in \mathbb{N}$, truncated to $m + n \leq 4$ which is enough for single photon characterization) which contains the information of the propagating mode signal quantum state from the histograms (See Appendix.~\ref{sec:app5}). Then, we derive Wigner function from signal moments and use maximum likelihood estimation (MLE) methods~\cite{smolin2012efficient} to determine the density matrix of the signal mode. The typical results of shaped photon states $(\ket{0}+\ket{1})/\sqrt{2}$ and $\ket{1}$ with frequency calibration by chirping are shown in Fig.~\ref{fig:4}. We then use the definition of fidelity of mixed quantum states ~\cite{jozsa1994fidelity} to calculate the fidelity of $(\ket{0}+\ket{1})/\sqrt{2}$ and $\ket{1}$ state, the fidelity is defined as
    \begin{align}
    F(\rho, \sigma) = \left( \text{Tr} \sqrt{\sqrt{\rho} \sigma \sqrt{\rho}} \right)^2,
    \end{align}
    where $\sigma$ is the measured density matrix and $\rho$ is the ideal density matrix. As a comparison, we also perform quantum state tomography of the shaped photon without frequency calibration by chirping. The results along with fidelity of the shaped photon with frequency calibration by chirping are all shown in Table.~\ref{tab:2}. The fidelity of $1/\sqrt{2}\ket{0}+\ket{1}$ state is $94.51 \pm 0.27\%$ and $94.24 \pm 0.06\%$, respectively. The fidelity of $\ket{1}$ state is $86.24 \pm 0.29 \%$ and $85.64 \pm 0.17 \%$, respectively. The fidelity of $\ket{1}$ state represents the qubit-photon energy exchange efficiency, in other words, the emission efficiency of our device and protocol. We observe that the fidelity of the shaped photon remains largely unaffected in the absence of frequency calibration. This can be attributed to the slight photon phase accumulation caused by the small amplitude of the parametric modulation, aligning with the photon time-dependent phase results illustrated in Fig.~\ref{fig:2}(d),(e). The primary cause of infidelity is the limited internal quality factor of the emission resonator and the limited relaxation time of the qubit which mainly results from the Purcell effect induced by the emission resonator. It can be mitigated by Purcell filter design~\cite{houck2008controlling,reed2010fast,jeffrey2014fast,bronn2015broadband,chen2023transmon} and fabrication improvement~\cite{ganjam2023surpassing,goetz2016loss}. 

    \begin{table}[htbp]
        \centering
        \renewcommand{\arraystretch}{1.5}
        \begin{tabular}{|c|c|c|}
            \hline
            \textbf{States} & $(\ket{0}+\ket{1})/\sqrt{2}$ &  $\ket{1}$ \\ \hline
            \textbf{With freq. calibration} &  $94.51 \pm 0.27\%$ &  $86.24 \pm 0.29 \%$  \\  \hline
            \textbf{Without freq. calibration}  & $94.24 \pm 0.06\%$  &  $85.64 \pm 0.17 \%$ \\ \hline
        \end{tabular}
        \caption{\textbf{Fidelity of typical states of shaped photons measured by photon quantum state tomography.}}
        \label{tab:2}
    \end{table}
    
\begin{figure}[htbp]
        \centering
        \includegraphics[width=0.45\textwidth]{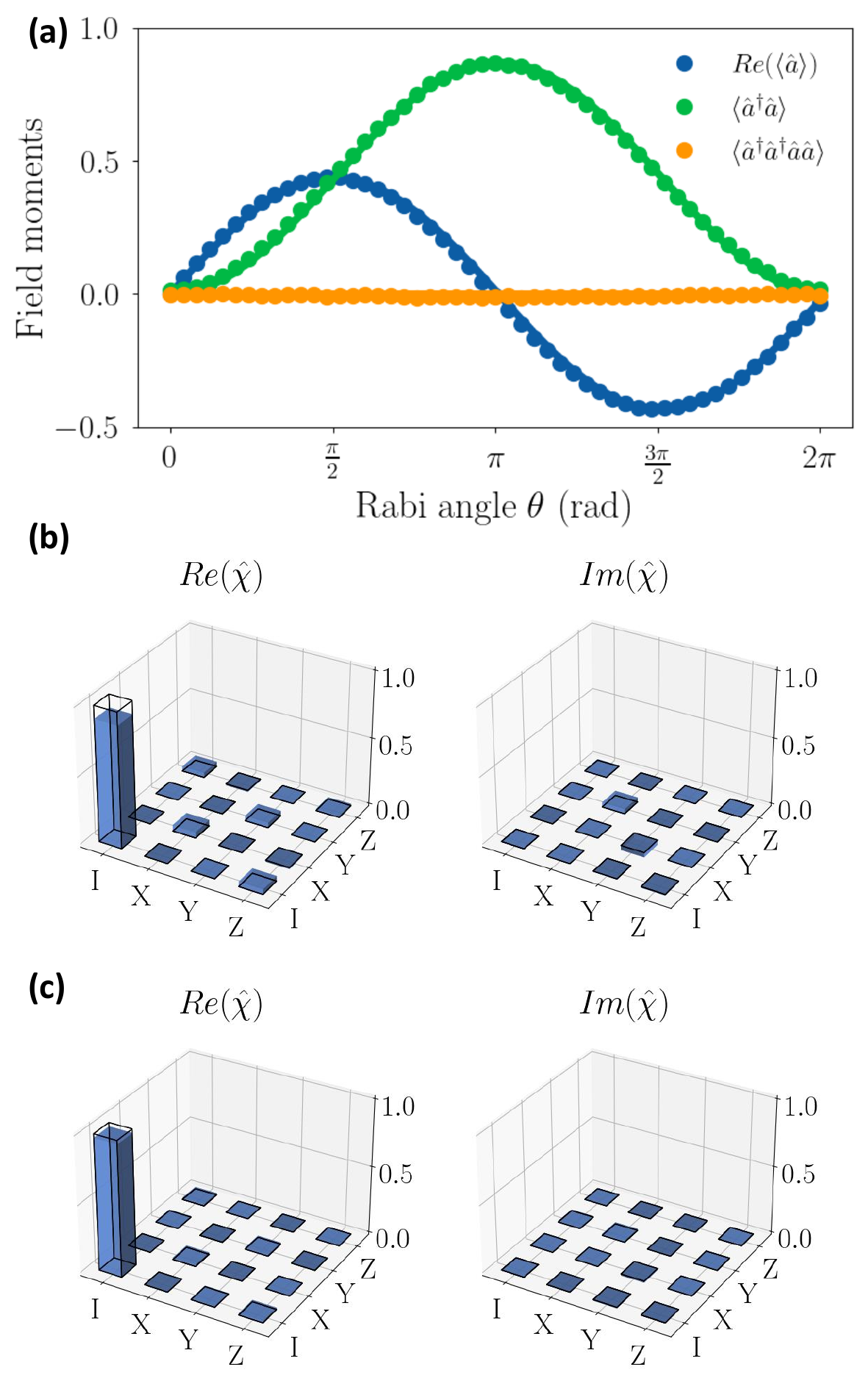}
        \caption{\textbf{Typical moments of shaped photon states and quantum process tomography of quantum state transfer}. \textbf{(a)}The first order, second order and fourth order moments of different Rabi angles. The dot is the experimental data averaging for $3\times10^8$ repeats and the solid line is the theoretical prediction considering the limited emission efficiency. \textbf{(b)}$\chi$ matrix of quantum state transfer using shaped photon derived by quantum process tomography. \textbf{(c)}$\chi$ matrix of quantum state transfer using unshaped photon derived by quantum process tomography.}
        \label{fig:5}
    \end{figure}

    Then by measuring the first order $\braket{\hat{a}}$, second order $\braket{\hat{a}^{\dagger}\hat{a}}$ and fourth order $\braket{\hat{a}^\dagger\hat{a}^\dagger\hat{a}\hat{a}}$ moments of photon field of different Rabi angles $\theta$, as shown in Fig.~\ref{fig:5}(a) (we only show $Re(\braket{\hat{a}})$ because $Im(\braket{\hat{a}})\approx0$), we confirm our shaped photon emission is in the single photon subspace. We obtain that the second-order correlation function $g^{(2)}(0)=\braket{\hat{a}^\dagger\hat{a}^\dagger\hat{a}\hat{a}}/(\braket{\hat{a}^{\dagger}\hat{a}})^{2}$ is $-0.060 \pm 0.035 $ and $-0.035 \pm 0.007$ for $1/\sqrt{2}(\ket{0}+\ket{1})$ and $\ket{1}$ state respectively which is a significant signal of photon anti-bunching and sub-Poissonian photon statistics~\cite{gu2017microwave}. Notified that our results of the fourth-order moment $\braket{\hat{a}^\dagger\hat{a}^\dagger\hat{a}\hat{a}}$ is all near zero but sightly negative as Ref.~\cite{Reuer2022realization}.

    Finally, we prepare six mutually-unbiased states to perform quantum process tomography (QPT)~\cite{nielsen2010quantum,poyatos1997complete} verifying the quality of quantum state transfer of our device and protocol. For each result shown in Fig.~\ref{fig:5}(b-c) we repeat for $3\times 10^8$ times and derive the $\chi$ matrix of quantum state transfer process, we define $\chi$ matrix as follows
    \begin{align}
    \mathcal{E}(\rho) = \sum_{m,n} \chi_{mn} A_m \rho A_n^\dagger,
    \end{align}
    where $\mathcal{E}$ represents a quantum operation, {$A_m$} is the two-dimensional complete operator base set $\{I,\sigma_{x},\sigma_{y},\sigma_{z}\}$, $\chi_{mn}$ is the matrix element of $\chi$ matrix. We recall the definition of the quantum process tomography fidelity by
    \begin{align}
    F(\chi_{\textcolor{red}{\text{ideal}}}, \chi) = \left( \text{Tr} \sqrt{\sqrt{\chi_{\textcolor{red}{\text{ideal}}}} \chi \sqrt{\chi_{\textcolor{red}{\text{ideal}}}}} \right)^2,
    \end{align}
    where $\chi$ is the measured process matrix, $\chi_{\textcolor{red}{\text{ideal}}}$ is the ideal process matrix. The fidelity of quantum state transfer based on shaped photon is 90.32\% and unshaped photon is 97.20\% as a comparison. The QPT fidelity of unshaped photon is higher than that of shaped photon. Since unshaped photons are produced by a \SI{40}{ns} flattop pulse and shaped photons are generated by a \SI{400}{ns} $\sin^2$ pulse, the qubit experiences more significant decay during the generation of shaped photons. This results further illustrate that the limited energy relaxation time of qubit is the main limitation of the emitted photon fidelity.
        
\section{Protocols for quantum state transfer and remote entanglement generation}

\label{sec:6}
\begin{figure}[htbp]
        \centering
        \includegraphics[width=0.45\textwidth]{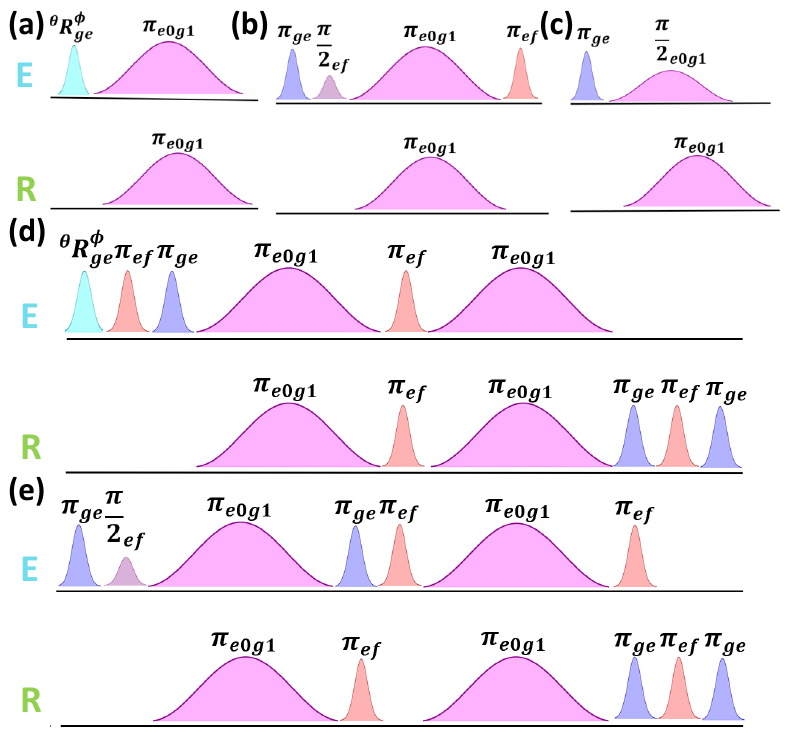}
        \caption{\textbf{Control pulse sequence protocols for emission nodes and receiving nodes of quantum state transfer and remote entanglement generation based on Fock state encoding and time-bin encoding.} E represents the emission node, R represents the receiving node. \textbf{(a)}Control sequence of Fock state encoded quatnum state transfer. \textbf{(b-c)}Control sequence of Fock state encoded remote entanglement generation with/without $\pi/2_{e0g1}$ pulse. \textbf{(d)}Control sequence of time-bin encoded quantum state transfer. \textbf{(e)}Control sequence of time-bin encoded remote entanglement generation.}
        \label{fig:6}
    \end{figure}

    To fully realize a point-to-point quantum network, we design a receiving protocol matched with the single-rail and dual-rail photon emission protocol (as shown in Fig.~\ref{fig:6}). For single-rail encoding quantum state transfer, we prepare a state $\alpha\ket{g}+\beta\ket{e}$ then use $\pi_{e0g1}$ pulses to pitch and catch (Fig.~\ref{fig:6}(a)) the shaped photon. There are two choices of single-rail remote entanglement generation protocol (Fig.~\ref{fig:6}(b-c)): the first one generates qubit-photon entanglement with the help of $\ket{f}$ state but does not require $\pi/2_{e0g1}$ pulse which brings about the process $\ket{e0} \to 1/\sqrt{2}(\ket{e0}+\ket{g1})$ and can use the $\pi_{e0g1}$ pulse calibrated in quantum state transfer process without much revise in the experiment; the second one does not require the $\ket{f}$ state instead uses $\pi/2_{e0g1}$ pulse to generate qubit-photon entanglement to perform 'half-pitch and full-catch' process in this case we must calibrate $\pi/2_{e0g1}$ pulse at emission node and re-calibrate $\pi_{e0g1}$ pulse at receiving node. In contrast to Ref.~\cite{pechal2014microwave,kurpiers2018deterministic,magnard2020microwave,storz2023loophole}, our protocol can avoid the use of $\ket{f}$ state. This is essential for us to obtain a higher transfer efficiency because energy relaxation time of $\ket{f}$ state $T_{1ef}$ is lower than that of $\ket{e}$ state $T_{1ge}$.
    
    For dual-rail time-bin photon, we must utilize $\ket{f}$ state as an ancillary state in both emission node and receiving node. For time-bin encoding quantum state transfer, we generate time-bin photon at emission node and then catch them one by one (Fig.~\ref{fig:6}(d)). For time-bin encoding remote entanglement generation, we first generate entanglement between the qubit and two time-bin mode then catch the time-bin mode one by one at receiving node (Fig.~\ref{fig:6}(e)). The time-bin photon is robust against photon transfer loss~\cite{kurpiers2019quantum} (with the cost of repetition rate) and phase reference error between sender and receiver~\cite{ilves2020demand}. Therefore, it is particularly useful for schemes using commercial circulators to ensure the communication channel unidirectionality and avoid the back-action of reflected photon which also causes unavoidable photon transfer loss.

    \textcolor{red}{Using the quantum state tomography data in Sec.~\ref{sec:5},  the transfer efficiency and the fidelity of quantum state transfer process~\cite{sete2015robust} can also be estimated assuming that the receiver setup has the same specifications. Because the $\ket{1}$ state fidelity is $F_{\ket{1}} = 86.24\pm0.29\%$ which also represents the energy emitting efficiency. Considering the transfer loss (including PCB, SMP adapters, circulator, low loss waveguide etc., 6\% in total.)~\cite{magnard2020microwave,storz2023loophole} and the energy absorption inefficiency (estimated to be $F_{\ket{1}}$),we can get the transfer efficiency $\eta_t = 0.86^2 \times (1-0.06) \approx 0.70$ provided perfect frequency matching between the two buffer resonators. Furthermore, the fidelity of quantum state transfer process can be estimated to be $F_{\chi} = (1+\sqrt{\eta_t})^2/4 \approx 0.84$. These results are at the same level as Ref.~\cite{kurpiers2018deterministic,magnard2020microwave,storz2023loophole}.}
        
\section{Discussion and Conclusion}
    To summarize, our study presents an on-demand shaped photon generation method utilizing a parametrically modulated qubit, achieving a high quantum process tomography fidelity of 90.32\% for single-rail shaped photon emissions, all without the need of additional tunable couplers. We implement slight parametric modulation pulses and use the first-order side-band coupling between the qubit and emission resonator inducing a linear relationship between parametric modulation amplitude and effective coupling covering a large tunable range up to \SI{20}{MHz}. This makes photon shaping and frequency calibration process experimentally friendly. Although the parametric modulation also results in a spurious frequency shift of qubit, for the optimal shaped photon, the phase accumulation of photon caused by the shift is almost negligible and we can further calibrate the dynamical frequency shift by chirping. Furthermore, our method is robust to cross-talk~\cite{zhou2021rapid}, enabling simultaneous execution of single qubit operations and the emission and reception of shaped photons across multiple channels on a multi-qubit chip. Therefore, our approach avoids the necessity for high-power control pulses~\cite{pechal2014microwave,kurpiers2018deterministic,kurpiers2019quantum}, enables easy amplitude and phase symmetrical photon generation and holds promise for scalable applications.
    
    In our case, the limitation of shaped photon emission speed is the decay rate of emission resonator. To speed up the emission process, we need to enhance the decay rate $\kappa_{E}/2\pi$ of our emission resonator up to 10-20 MHz. To protect the qubit from Purcell decay,
    Purcell filters~\cite{houck2008controlling,reed2010fast,jeffrey2014fast,bronn2015broadband,chen2023transmon} need to be added to the design. By enhancing the qubit energy relaxation time and speeding up the emission process, we can reach a higher emission efficiency and quantum process tomography fidelity of quantum state transfer.
    
    For future improvements in experimental techniques, we can use absolute power calibration methods like measurement-induced dephasing~\cite{kono2018quantum} and AC Stark shift induced by emission resonator to get a more accurate gain of the amplification chain and absolute value of experimental data. For pulse shaping, we should further consider the transfer function of our flux line and calibrate the distortion of parametric modulation pulses and use numerical methods to optimize them.

    Further using the parametric modulation side-band coupling between higher state of qubit and emission resonator(the first-order side-band coupling between $\ket{f0}$ and $\ket{e1}$),  we can generate complex 1D time-bin photon entangled states~\cite{besse2020realizing}. With the help of the emission resonator we can also implement quantum non-demolition detection of itinerant photon~\cite{kono2018quantum,besse2018single}, parity detection~\cite{besse2020parity}, universal quantum gate set for itinerant photons~\cite{Reuer2022realization} and photon scattering gate like Duan-Kimble gate~\cite{duan2004scalable} for remote controlled-phase gate between quantum nodes~\cite{penas2022universal}.

    
\section*{data availability statement}
The data produced in this work is available from the corresponding authors upon reasonable request.

\begin{acknowledgments}
The authors thank Prof. Hongyi Zhang and Yan Li from Tsinghua University and Prof. Zeliang Xiang from Sun Yat-sen University for insightful discussions. This work was supported by the Micro/Nano Fabrication Laboratory of Synergetic Extreme Condition User Facility
(SECUF). Devices were made at the Nanofabrication Facilities at the Institute of Physics, CAS in Beijing. The authors thank Beijing Naishu Electronics Co., Ltd. for providing support of RF-DAC and RF-ADC based on RFSoC FPGA. This work was supported by: Innovation Program for Quantum Science and Technology (Grant No. 2021ZD0301800), Strategic Priority Research Program of Chinese Academy of Sciences (Grant No. XDB28000000), the National Natural Science Foundation of China (Grants No.12204528, 92265207, T2121001, 12005155, 92065114, 12375024 and 12047502), Scientific Instrument Developing Project of Chinese Academy of Sciences (Grant No. YJKYYQ20200041) and Beijing Natural Science Foundation (Grant No. Z200009).
\end{acknowledgments}

\section*{author contributions}
D.-N.Z., Y.-X.Z. and Z.-C.X. supervised the project. X.L. conceived the idea, designed the sample, performed the experiment, processed the data and wrote this manuscript. X.L. and S.-Y.L. built up the experiment measurement and control system with RFSoC and the data stream processing of GPU. S.-L.Z. fabricated the sample. Z.-Y.M. and X.-H.S. fabricated the JPA. D.-N.Z. and Y.-X.Z. revised the manuscript. All authors contributed to the discussions and production of the manuscript. 

\section*{competing interests}
The authors declare no competing interests.

\appendix
\renewcommand{\thefigure}{S\arabic{figure}}
\setcounter{figure}{0}
\renewcommand{\thetable}{S\arabic{table}}
\setcounter{table}{0}
\section{\MakeUppercase{Sample fabrications and sample parameters}}
\label{sec:app1}
    \begin{figure}[htbp]
        \centering
        \includegraphics[width=0.45\textwidth]{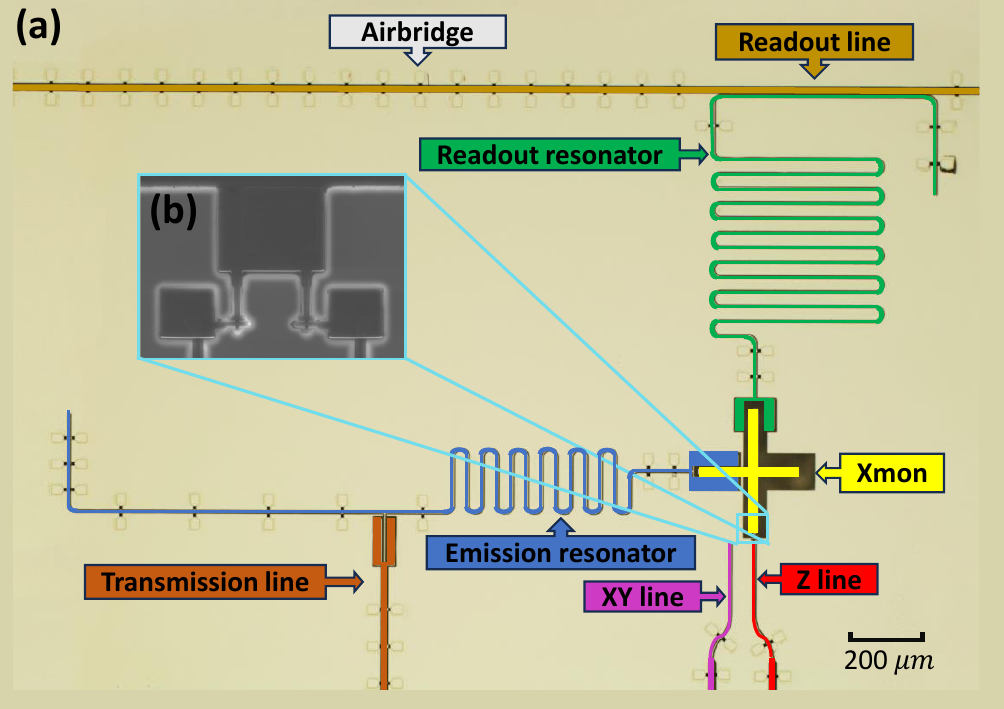}
        \caption{\textbf{Sample optical microscope diagram with false colors.}\textbf{(a)}The main parts of the sample layout where there are one Xmon capacitively coupled to readout resonator and emission resonator. An individual XY line and Z line is to drive the qubit and change the flux of the DC-SQUID loop. We fabricated air-bridge at each CPW resonators and transmission lines to prevent the propagation of parasitic slotline modes\cite{chen2014fabrication}. \textbf{(b)}The SEM image of DC-SQUID loop which is connected to the ground.}
        \label{fig:a1}
    \end{figure}
    
    The device is fabricated using micro-nano fabrication technology, the main steps are as follows: (1)Growth and patterning of aluminum film. A \SI{100}{nm} aluminum thin film is uniformly grown on a c-plane sapphire substrate by electron beam evaporation (\text{Plassys-MBE550s}). Shunted-capacitors, resonators and transmission lines are patterned by laser direct writing (\text{DWL66+}) and etched by wet etching (\text{ZX238}). (2)Preparation of josephson junction. Josephson junctions are formed by crossing two perpendicular lines using double-angle electron beam evaporation of aluminum through a Dolan-bridge prepared by spin-coating double-layer photo-resist and electron beam exposure (\text{Raith EBPG5200}). (3)Preparation of Airbridge. The use of airbridges in device ensures good ground plane connectivity and reduces slotline modes~\cite{chen2014fabrication}. \SI{300}{nm} airbridges are prepared by laser direct writing (\text{DWL66+}) and wet etching (\text{Aluminum Etchant Type D}). (4)Slicing and Packaging. The device is cut into 6×6 \si{\square\milli\meter} (\text{DISCO-DAD323}), wire-bonded (\text{Westbond-7476d}) to a silver-plated printed circuit board (PCB) and then packed in a aluminum packaging box.

    For sample design, we engineer a large coupling (about \SI{300}{MHz}) between qubit and emission resonator to have a large tunable effective coupling range where effective coupling is linear to parametric modulation amplitude. The emission resonator decay rate is designed to be large (about \SI{5.2}{MHz}) to speed up the photon emission process. The measured sample parameters are shown in Table.~\ref{tab:a1}.

    \begin{table}[htbp]
        \centering
        \begin{tabular}{|c|c|}
            \hline
            \textbf{Parameters} &  \\ \hline
            Qubit Max Frequency $\omega_{q,p}/2\pi$ & 5.997 GHz \\  \hline
            Qubit Operating Frequency $\omega_{q}/2\pi$ &  5.947 GHz \\ \hline
            Qubit Anharmonicity $\alpha/2\pi$ & -228 MHz \\ \hline
            Qubit Energy Relaxation Time for $\ket{e}$ $T_{1ge}$ &  5.539 $\mu s$ \\ \hline
            Qubit Energy Relaxation Time for $\ket{f}$ $T_{1ef}$ &  2.829 $\mu s$ \\ \hline
            Qubit Decoherence Time for $\ket{e}$ $T^{*}_{2ge}$ &  2.234 $\mu s$ \\ \hline
            Qubit Decoherence Time for $\ket{f}$ $T^{*}_{2ef}$ &  1.068 $\mu s$ \\  \hline
            Emission Resonator Frequency $\omega_{E}/2\pi$ &  7.139 GHz\\ \hline
            \textcolor{red}{Emission Resonator Coupling Decay Rate $\kappa_{E,c}/2\pi$} &  5.2 MHz \\  \hline
            \textcolor{red}{Emission Resonator Internal Decay Rate $\kappa_{E,i}/2\pi$} & \textcolor{red}{0.27 MHz} \\ \hline
            Emission Resonator-Qubit Coupling $g_{qE}/2\pi$ & 299.4 MHz  \\ \hline
            Readout Resonator Frequency $\omega_{R}/2\pi$ &  4.177 GHz \\ \hline
            Readout Resonator Decay Rate $\kappa_{R}/2\pi$ &  0.4 MHz \\ \hline
            Readout Resonator-Qubit Coupling $g_{qR}/2\pi$ &  48.2 MHz \\
            \hline
        \end{tabular}
        \caption{\textbf{Summary of sample parameters.}}
        \label{tab:a1}
    \end{table}

\section{\MakeUppercase{Experimental setup}}
\label{sec:app2}
    \begin{figure}[htbp]
        \centering
        \includegraphics[width=0.45\textwidth]{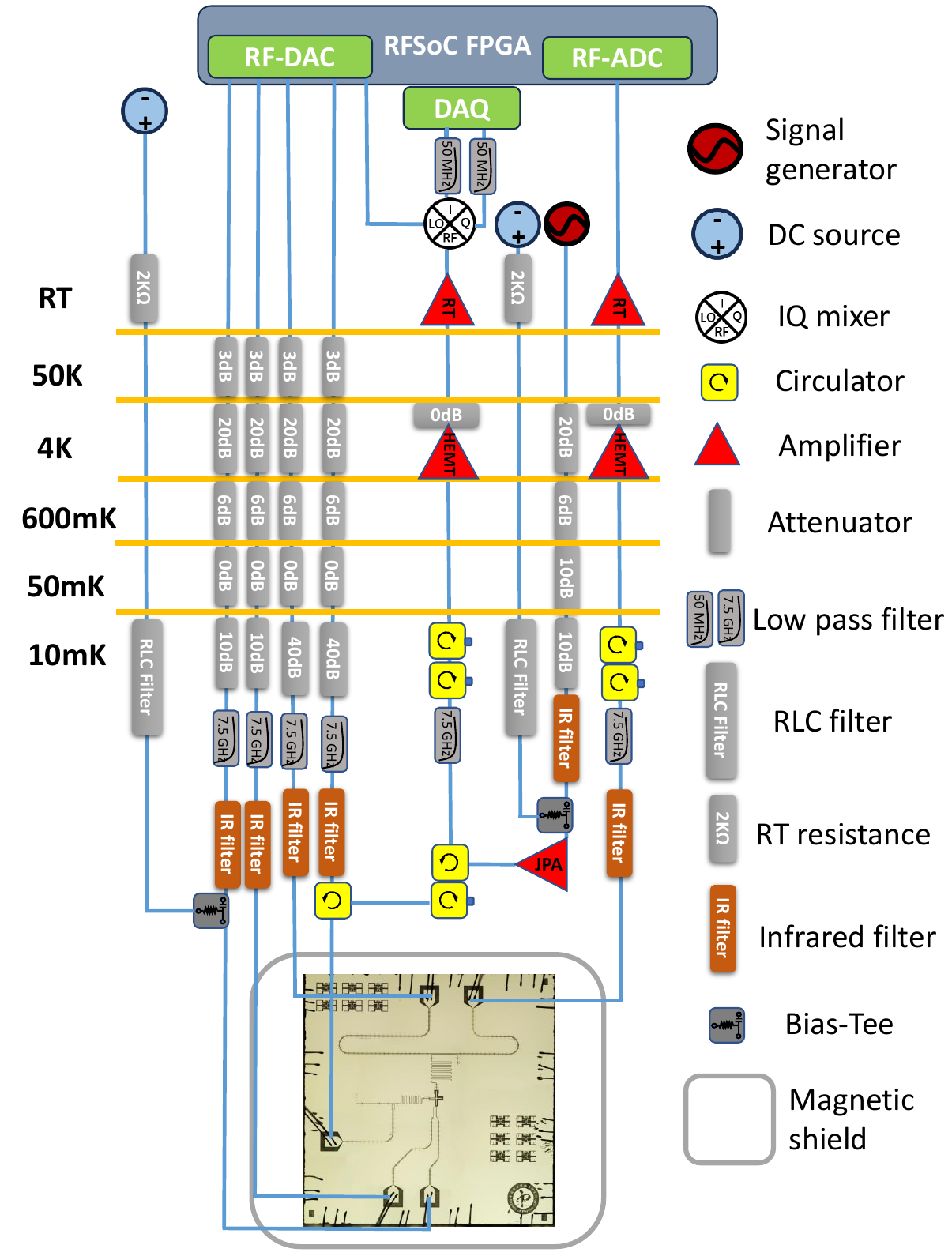}
        \caption{\textbf{The experimental setup in dilution refrigerator and room temperature.}}
        \label{fig:a2}
    \end{figure}
    We mount our sample inside a magnetic shield in the mixing chamber stage of a Bluefors dilution refrigerator. We use eccosorb infrared filters~\cite{corcoles2011protecting} in every RF microwave input coaxial line to mitigate quasi-particle excitation, \SI{7.5}{GHz} low pass filters to mitigate higher harmonics and proper attenuators to mitigate the Johnson-Nyquist noise from room temperature. The XY line and Z line of qubit are separated and have the same attenuator and filter arrangement. The Z line is combined with a DC line which has a RLC filter and provided stable DC flux bias to qubit by a bias-tee. We have two read-in-and-out loop, one is for qubit dispersive readout, the other is for the characterization of emission resonator and measurement of photon field. Thus, we can measure the $S_{21}$ of the readout resonator and the $S_{11}$ of the emission resonator. A flux-driven impedance-engineered JPA~\cite{yamamoto2008flux,roy2015broadband} which has a flux-pump line and DC bias line combined by a bias-tee is used to enhance the measurement quantum efficiency for photon quadrature measurement and tomography. An extra circulator is put between the sample and JPA to mitigate the measurement back-action of JPA. In the two read-out line, there are both high-electron-mobility transistor (HEMT) and room temperature (RT) amplifier aimed at amplifying the signal to proper voltage for analog-digital converter (ADC) to sample.
    
    For electronics in room temperature, we build up a control and measurement system with radio-frequency system-on-chip (RFSoC)~\cite{tholen2022measurement,park2022icarus,stefanazzi2022qick} to directly generate and sample radio-frequency signal for qubit control and readout. We use a radio-frequency digital-analog converter (RF-DAC) based on RFSoC FPGA (\text{Xilinx Zynq UltraScale+ RFSoC XCZU47DR}) whose sampling rate is \SI{8.0}{GSa/s} to directly generate read-in and XY control pulses without modulation using the second Nyquist zone with the help of extra band pass filters. The generated frequency range is 4.0-8.0 \si{GHz} and suitable for qubit measurement and control. The mix-mode~\cite{park2022icarus,Kalfus2020high} is used to have a smooth power-frequency dependence in higher Nyquist zones. The parametric modulation pulse is generated by the same RF-DAC using the first Nyquist zone with the help of a \SI{4.0}{GHz} low pass filter. For qubit readout signal sampling, we use a radio-frequency analog-digital converter (RF-ADC) with \SI{4.0}{GSa/s} sampling rate based on another RFSoC FPGA (\text{Xilinx Zynq UltraScale+ RFSoC XCZU47DR}) to directly sample signals at higher Nyquist zones which are aliased within the first Nyquist zone of RF-ADC. To avoid signal aliasing from other Nyquist zones and enhance signal-to-noise ratio (SNR) we also add band pass filter in front of the RF-ADC.

    In order to meet the demands of massive data processing for low SNR experiments like microwave photon field heterodyne measurement and tomography, we build a data stream processing system with a NVIDIA GPU and a high speed data acquisition (DAQ) card with \SI{1.0}{GSa/s} sampling rate. We are able to process the experimental data in real-time, eliminating the need for storage, which in turn reduces the time cost. In our experiment, the trigger period is set to \SI{12}{\micro\second}, and the experiment repeated $3 \times 10^8$ times only cost 1 hour in total. The data processing and transfer time between the DAQ card and the GPU is not the limitation. We can further improve the duty cycle and reduce the experimental time cost by reducing the trigger period because photon emission is just qubit reset. Furthermore, we utilize the output of a single RF-DAC channel, augmented with a room temperature amplifier, to serve as the local oscillators (LO) for the heterodyne measurements, guaranteeing phase locking for each measurement. All the electronic devices in room temperature are synchronized by a rubidium clock (Stanford Research Systems FS725) and triggered by the same clock and trigger distribution module.

\section{\MakeUppercase{Detailed calculations about parametric modulation}}
\label{sec:app3}
    It is convenient and reasonable to omit the readout resonator in our flux modulating process, then consider a two level system approximation, we will derive a Jaynes-Cummings like Hamiltonian 
    
    \begin{align}
    \hat{H}/\hbar = \frac{\omega_q(\Phi(t))}{2} \hat{\sigma}_z + \omega_E \hat{a}^\dagger \hat{a} + g_{qE}(\hat{a}^\dagger \hat{\sigma}_- + \hat{a} \hat{\sigma}_+),
    \end{align}
    where $\hat{a}$($\hat{a}^{\dagger}$) is the annihilation(creation) operator of the emission resonator mode, $\hat{\sigma}_-$($\hat{\sigma}_+$) is the annihilation(creation) operator of qubit \{$\ket{g}$,$\ket{e}$\} subspace. Because of the non-linearity of frequency-flux relation of transmon, when we add sinusoidal parametric modulation $\Phi(t)=\Phi_{\text{DC}}+\Phi_{\text{AC}}\cos(\omega_{m}t+\theta_{m})$ there are high order harmonics terms $A^a_m\cos[a(\omega_{m}t+\theta_{m})](a \in \mathbb{N})$ and DC offset $\Delta_{DC}$ in qubit frequency. We call the $a$th order harmonics term of qubit time-dependent frequency variation the $a$th order longitudinal field modulation (LFM). When qubit is not at the sweet spot, the least order of longitudinal field modulation (LFM) is $a=1$, and we ignore the higher order terms, then we have
    \begin{align}
    \omega_q(\Phi(t)) &= \overline{\omega}_q + \sum_{a=1}^{\infty} A^{a}_{m}\cos[a(\omega_{m}t+\theta_{m})] \nonumber \\
    &\approx \overline{\omega}_q + A^{1}_{m}\cos(\omega_{m}t+\theta_{m}),
    \end{align}
    where
    \begin{align}
    \overline{\omega}_q=\omega_q(\Phi_{\text{DC}}) + \Delta_{\text{DC}}(\Phi_{\text{DC}},\Phi_{\text{AC}}),
    \end{align}
    \begin{align}
    \label{eq:c5}
    A^{1}_{m}(\Phi_{\text{DC}},\Phi_{\text{AC}}) = \Phi_{\text{AC}}\left.\frac{d\omega_q}{d\Phi}\right|_{\Phi_{\text{DC}}},
    \end{align}
    \begin{align}
    \Delta_{\text{DC}}(\Phi_{\text{DC}},\Phi_{\text{AC}}) = \frac{\Phi^{2}_{\text{AC}}}{4} \left.\frac{d^2\omega_q}{d\Phi^{2}} \right|_{\Phi_{\text{DC}}}.
    \end{align}
    Then we know that the first order of LFM is proportional to the parametric modulation amplitude $A^{1}_{m} \propto \Phi_{AC}$ and there is a quadratic relation between $\Delta_{\text{DC}}$ and $\Phi_{\text{AC}}$. Both LFM amplitude $A^{1}_{m}$ and DC offset $\Delta_{\text{DC}}$ are dependent on the working point $\Phi_{\text{DC}}$ and parametric modulation amplitude $\Phi_{\text{AC}}$. When the $\Phi_{\text{DC}}$ is larger, $A^{1}_{m}$ is more sensitive to $\Phi_{\text{AC}}$ while $\Delta_{\text{DC}}$ is less sensitive to $\Phi_{\text{AC}}$ according to the frequency-flux relation of transmon with symmetric Josephson junction SQUID~\cite{koch2007charge}. In our experiment, we choose $\Phi_{\text{DC}}=0.04$ to balance the trade-off of LFM frequency shift, tunable coupling range and qubit coherence.
    
    We use a time-dependent unitary transformation 
    \begin{align}
    &\hat{U}(t) = \nonumber \\
    &\exp\left\{-i \left(\frac{\overline{\omega}_q}{2} \hat{\sigma}_z + \frac{A^1_m}{2\omega_m} \cos(\omega_m t + \theta_m) \hat{\sigma}_z + \omega_E \hat{a}^\dagger \hat{a}\right)\right\},
    \end{align}
    to show side-band couplings in a rotating frame, in the rotating frame the Hamiltonian is 
    \begin{align}
    &\hat{H}'/\hbar = \hat{U}\hat{H}/\hbar\hat{U}^\dagger - i\hat{U}\partial_t\hat{U}^\dagger \nonumber \\
    &= g_{qE} e^{-i\Delta_{qE} t} \hat{a}^\dagger \hat{\sigma}_- \sum_{n=0}^{\infty}  J_n\left( \frac{A^1_m}{\omega_m} \right) e^{i (n\omega_m t + \theta_m +\frac{\pi}{2})}  + h.c.,
    \end{align}
    where $\Delta_{qE} = \overline{\omega}_q - \omega_E$ \textcolor{red}{and $J_n(x)$ is the nth order Bessel function}. When $A^1_m$ is small ($\Phi_{\text{AC}}$ is small) and $\Delta_{qE} \approx \omega_m$, only the zero and first-order side-bands have a significant impact.
    
    The zero side-band coupling with the emission resonator is a dispersive coupling 
    which induces Lamb shift to both qubit frequency $\tilde{\omega}_q = \omega_q + \Delta_{\text{Lamb}}$ and emission resonator $\tilde{\omega}_E = \omega_E - \Delta_{\text{Lamb}}$,
    \begin{align}
    \Delta_{\text{Lamb}} &= \frac{\left(g_{qE} J_0\left(\frac{A^1_m}{\omega_m}\right)\right)^2}{\Delta_{qE}} \nonumber \\
    &= \frac{\left(g_{qE} \left(1-\frac{(A^1_m)^2}{4\omega^2_m}\right)\right)^2}{\Delta_{qE}} + O\left( \left(\frac{A^1_m}{\omega^2_m}\right)^4\right)\nonumber \\
    &= \frac{g_{qE}^2}{\Delta_{qE}} \left(1-\frac{(A^1_m)^2}{2\omega^2_m}\right) + O\left(\left(\frac{A^1_m}{\omega^2_m}\right)^4\right),
    \end{align}
    \textcolor{red}{where $J_0(x)$ represents the zeroth order Bessel function (when $x \to 0$, $J_0(x) \approx 1 - x^2/4$)}. Then, the zero-order side-band induced Lamb shift variation is
    \begin{align}
    \Delta'_{\text{Lamb}} &= 2\left(\Delta_{\text{Lamb}} - \frac{g_{qE}^2}{\Delta_{qE}} \right) \nonumber \\
    & = -\frac{g_{qE}^2}{\Delta_{qE}}\frac{(A^1_m)^2}{\omega^2_m},
    \end{align}
    which represents the zero-order side-band induced dressed frequency variation between the qubit and emission resonator and has a quadratic relation of $\Phi_{\text{AC}}$. In our scenario, we estimate the Lamb shift to be \SI{3.0}{MHz} at $\Phi_{\text{AC}}=0.25\Phi_0$, which is an order of magnitude smaller than $\Delta_{\text{DC}}$ (about \SI{-40}{MHz}, see Appendix.~\ref{sec:app4}), though, still produces an observable effect. However, it becomes nearly negligible at $\Phi_{\text{AC}}=0.037$. It is worth noting that when the frequency of emission resonator is higher than the qubit, the sign of DC offset $\Delta_{\text{DC}}$ and Lamb shift $\Delta'_{\text{Lamb}}$ are the opposite and can compensate with each other. Generally, the DC offset induced by non-linearity of frequency-flux relation of transmon and Lamb shift induced by the zero-order side-band coupling are both quadratic to the parametric modulation amplitude $\Phi_{\text{AC}}$ and can be calibrated together.
    
    Considering the Lamb shift caused by the zero-order side-band, then when the first-order side-band of qubit and emission resonator are on resonance $\tilde{\omega}_E = \tilde{\omega}_q + \Delta_{\text{DC}} + \omega_m$ the Hamiltonian can be simplified to
    \begin{align}
    \hat{H}'/\hbar&=g_{qE}J_{1}(\frac{A^{1}_{m}}{\omega_{m}})e^{i\theta_{m}}\hat{a}\hat{\sigma}_{+} + h.c. \nonumber \\
    &\approx g_{qE}\frac{A^{1}_{m}}{2\omega_{m}}e^{i\theta_{m}} \hat{a}\hat{\sigma}_{+} + h.c..
    \end{align}
    Finally, we get the effective coupling which is linear to the parametric modulation amplitude and has a phase dependence to the parametric modulation phase
    \begin{align}
    g_{\textcolor{red}{\text{eff}}} &= g_{qE} \frac{A^1_m}{2\omega_m}e^{i\theta_{m}} \nonumber \\
    &= g_{qE} \frac{\Phi_{\text{AC}}\left.\frac{d\omega_q}{d\Phi}\right|_{\Phi_{\text{DC}}}}{2\omega_m}e^{i\theta_{m}} \nonumber \\
    &\propto \Phi_{\text{AC}} e^{i\theta_{m}},
    \end{align}
    using Eq.~\ref{eq:c5}. Then we can use a time-dependent $\Phi_{AC}(t)$ to shape the photon temporal profile and time-dependent $\theta_{m}(t)$ to eliminate the frequency shift $\Delta(t)=\Delta_{\text{DC}}(t) + \Delta'_{\text{Lamb}}(t)$, where $\dot{\theta}_{m}(t)=-\Delta(t)$.

\section{\MakeUppercase{Calibrations of the first-order side-band parametric modulation}}
    \label{sec:app4}
    \begin{figure}[htbp]
        \centering
        \includegraphics[width=0.45\textwidth]{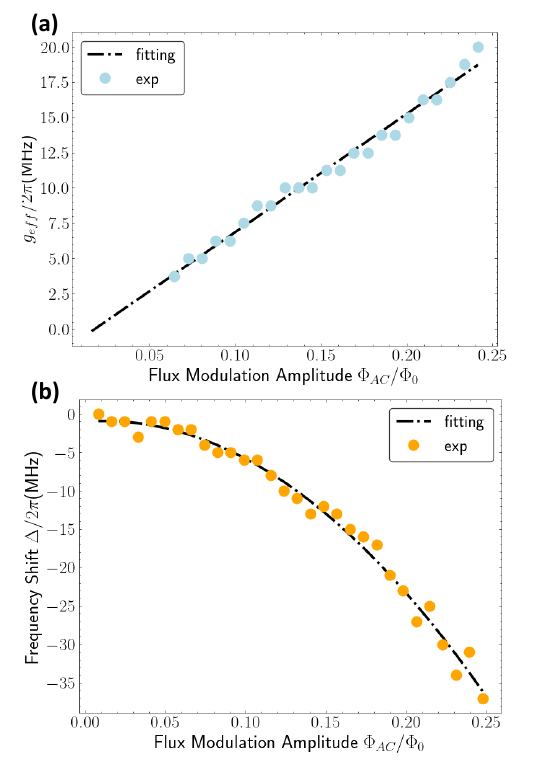}
        \caption{\textbf{Calibrations of the first-order side-band parametric modulation.} \textbf{(a)}The relation between the first-order side-band effective coupling and parametric modulation amplitude. The dots are experimental results and the dotted line is the linear fitting results. \textbf{(b)}The relation between the first-order side-band parametric modulation frequency shift and parametric modulation amplitude. The dots are experimental results and the dotted line is the quadratic fitting results.}
        \label{fig:a3}
    \end{figure}

    As Fig.~\ref{fig:a3}(a) shown, we calibrate the relation between parametric modulation amplitude $\Phi_{\text{AC}}$ and effective first-order side-band coupling $g_{eff}/2\pi$ by the flattop pulse(square pulse convoluted with Gaussian function with $2\sigma=4ns$ to have a smooth rising edge) shape with varying amplitudes and lengths. The effective coupling $g_{eff}/2\pi$ is greatly linear to parametric modulation amplitude $\Phi_{\text{AC}}$. The max effective coupling $g_{eff,max}/2\pi$ is about 20 MHz at $0.25\Phi_{0}$ parametric modulation amplitude where $\Phi_{0}$ is the magnetic flux quantum. The result shows that the first-order side-band coupling of parametric modulation in our design has a wide tunable range and can support fast photon emission. 
    
    As Fig.~\ref{fig:a3}(b) shown, we calibrate the parametric modulation frequency shift caused by average frequency shift and Lamb shift by the flattop pulse with varying amplitudes and frequencies. The frequency shift $\Delta = \Delta_{DC} + \Delta'_{\text{Lamb}}$ is quadratic to the parametric modulation amplitude $\Phi_{\text{AC}}$ and the max frequency shift is about -37 MHz at $0.25\Phi_{0}$ parametric modulation amplitude. For $\sin^2$ type photon shaping pulse used in our experiments, the amplitude is small($A_{\text{AC}}=0.037\Phi_{0}$) and the corresponding frequency shift is only about -0.4 MHz which causes almost negligible photon phase accumulation during the emission process.

\section{\MakeUppercase{Photon state tomography}}
\label{sec:app5}
    We use heterodyne detection method~\cite{Silva2010schemes} to measure the two quadratures of amplified single photon signal by a high speed data acquisition card. We use a template function $f(t)$($\int |f(t)|^2 \, dt=1$) that is the same shape as $(\ket{0}+\ket{1})/\sqrt{2}$ state quadratures as a matched filter to maximize the detection efficiency. Then we have a time-independent signal mode $\hat{a}=\int f(t)\braket{\hat{a}_{out}(t)} \, dt$. Considering the noise generated by phase-insensitive amplifier chain~\cite{caves1982quantum}, the final quadrature field we measured is 
    
    \begin{align}
    \hat{S}_0 = \sqrt{G}\hat{a} + \sqrt{G-1}\hat{h}^{\dagger},
    \end{align}
    where $\hat{a}$ is the an annihilation operator of signal mode, the $\hat{h}^{\dagger}$ is the creation operator of noise mode. When $G \gg 1$ then we have $\sqrt{G} \approx \sqrt{G-1}$, then we re-define the measured mode
    \begin{align}
    \hat{S} = \frac{\hat{S}_0}{\sqrt{G}} \approx \hat{a} + \hat{h}^{\dagger}.
    \end{align}
    We can generate a 2D histogram $D^{[\rho_a]}(S,S^{*})$ to describe quasi-probability distribution of measured quadratures. If we assume the $\hat{a}$ and $\hat{h}$ mode are uncorrelated, then the moments of $\hat{S}$, $\hat{a}$ and $\hat{h}^{\dagger}$ are related by

    \begin{align}
    \label{eq:e3}
    \braket{(\hat{S}^{\dagger})^n S^m}_{\rho_a} = \sum_{i,j=0}^{n,m} \binom{m}{j} \binom{n}{i} \braket{(\hat{a}^{\dagger})^i \hat{a}^j} \braket{\hat{h}^{n-i}(\hat{h}^{\dagger})^{m-j}}.
    \end{align}
    We can get the $\hat{S}$ mode moments from our 2D histogram $D^{[\rho_a]}(S,S^{*})$ 
    \begin{align}
    \braket{(\hat{S}^\dagger)^n S^m}_{\rho_a} = \int_{S} (S^*)^n S^m D^{[\rho_a]}(S,S^{*}) G^{-\frac{n+m}{2}}.
    \end{align}
    To obtain the moments of the noise added by the amplification chain $\braket{\hat{h}^{n}(\hat{h}^{\dagger})^{m}}$, we generate a 2D histogram of the background without photon emission $D^{\ket{0}\bra{0}}(S,S^{*})$ from which we extract the noise moments
    \begin{align}
    \label{eq:e5}
    \braket{\hat{h}^{n}(\hat{h}^{\dagger})^{m}} &= \braket{(\hat{S}^\dagger)^n S^m}_{\ket{0}\bra{0}} \nonumber \\
    &= \int_{S} (S^*)^n S^m D^{\ket{0}\bra{0}}(S,S^{*}) G^{-\frac{n+m}{2}}.
    \end{align}
     Then by inserting Eq.~\ref{eq:e5} to Eq.~\ref{eq:e3}, we get a linear equation of signal moments $\braket{(\hat{a}^{\dagger})^n \hat{a}^m}$. We obtain the signal moments from the linear equation then use the formula~\cite{haroche2006exploring,eichler2013experimental}
    \begin{align}
    W(\alpha) = \sum_{n,m} \int d^2 \lambda \frac{\braket{ (\hat{a}^\dagger)^n\hat{a}^m} (-\lambda^*)^m \lambda^n}{\pi^2 n! m!} e^{-\frac{1}{2}|\lambda|^2 + \alpha \lambda^* - \alpha^* \lambda}.
    \end{align}
    to derive Wigner function directly. The gain of amplification chain G used above is estimated by $\braket{\hat{a}^\dagger\hat{a}} = |\braket{\hat{a}}|$ relation of $(\ket{0}+\ket{1})/\sqrt{2}$ state. Then we use maximum-likelihood estimation(MLE) methods to get a physical density matrix $\rho_a$ truncated in single-photon subspace. We maximize the log-likelihood function~\cite{eichler2013experimental}
    \begin{align}
    \mathcal{L}_{\text{log}} = - \sum_{n,m} \frac{1}{\delta_{n,m}^2} \left|\braket{ (\hat{a}^\dagger)^n \hat{a}^m} - \text{Tr}[\rho_a (\hat{a}^\dagger)^n \hat{a}^m]\right|^2,
    \end{align}
    using convex optimization method~\cite{strandberg2022simple}(by CVXPY~\cite{diamond2016cvxpy}) to guarantee the semi-definite, Hermitian and trace one properties of density matrix, where $\delta_{n,m}$ is the standard deviation of signal moments $\braket{(\hat{a}^\dagger)^n \hat{a}^m}$.

    By extracting the moments of noise mode $\braket{\hat{h}^{n}(\hat{h}^{\dagger})^{m}}$, we get the average photon number of the thermal field of the amplification chain $n_{\text{noise}} = \braket{\hat{h}^{\dagger}\hat{h}} = 2.78$ and we then get the overall quantum efficiency which includes noise added by the amplifiers, loss and spurious reflections in the whole amplification chain and detection inefficiency of our sampling and data processing process~\cite{ferreira2024deterministic}
    \begin{align}
    \eta = \frac{1}{1+n_{\text{noise}}} \approx 0.26.
    \end{align}
    
    We only perform single mode photon field tomography in this experiment. Multi-mode tomography for time-bin mode and photon-qubit joint tomography characterizing the entanglement between them is straightforward by expanding the methods above~\cite{eichler2013experimental}.
    
\section{\MakeUppercase{Initial state quality for quantum state transfer}}
    We use $15 ns$ Gaussian pulse with DRAG~\cite{motzoi2009simple} to be the $\pi$ and $\pi/2$ pulse for both $\ket{g}$-$\ket{e}$ and $\ket{e}$-$\ket{f}$ state to optimize the single qubit gate fidelity and initial state preparation fidelity. We characterize the single qubit gate and initial state preparation fidelity by interleaved randomized benchmarking(IRB)~\cite{magesan2012efficient} and quantum state tomography(QST)~\cite{steffen2006measurement}. The results are shown in Table.~\ref{tab:a2}
    \begin{table} [htbp]
        \centering
        \begin{tabular}{|@{\hskip 0.2cm}c@{\hskip 0.2cm}|@{\hskip 0.5cm}c@{\hskip 0.5cm}|@{\hskip 0.5cm}c@{\hskip 0.5cm}|}
            \hline
            \textbf{Single qubit gates(states)} & \textbf{IRB} & \textbf{QST}\\ \hline
            I & 99.81\% & 99.92\% \\ \hline
            X & 99.75\% & 99.36\% \\ \hline
            Y/2 & 99.77\% & 99.89\% \\ \hline
            X/2 & 99.77\% & 99.90\% \\ \hline
            -Y/2 & 99.71\% & 99.53\% \\ \hline
            -X/2 & 99.74\% & 99.54\% \\
            \hline
        \end{tabular}
        \caption{\textbf{Single qubit gate fidelity by interleaved randomized benchmarking and initial state preparation fidelity by quantum state tomography.} I represents $\ket{g}$ state, X represents $\ket{e}$ state, Y/2 represents $(\ket{g}+\ket{e})/\sqrt{2}$ state, X/2 represents $(\ket{g}+i\ket{e})/\sqrt{2}$ state, -Y/2 represents $(\ket{g}-\ket{e})/\sqrt{2}$ state and -X/2 represents $(\ket{g}-i\ket{e})/\sqrt{2}$ state.}
        \label{tab:a2}
    \end{table}

\section{\MakeUppercase{Photon phase dependence of the first-order side-band parametric modulation}}
\label{sec:app6}
    \begin{figure}[htbp]
        \centering
        \includegraphics[width=0.45\textwidth]{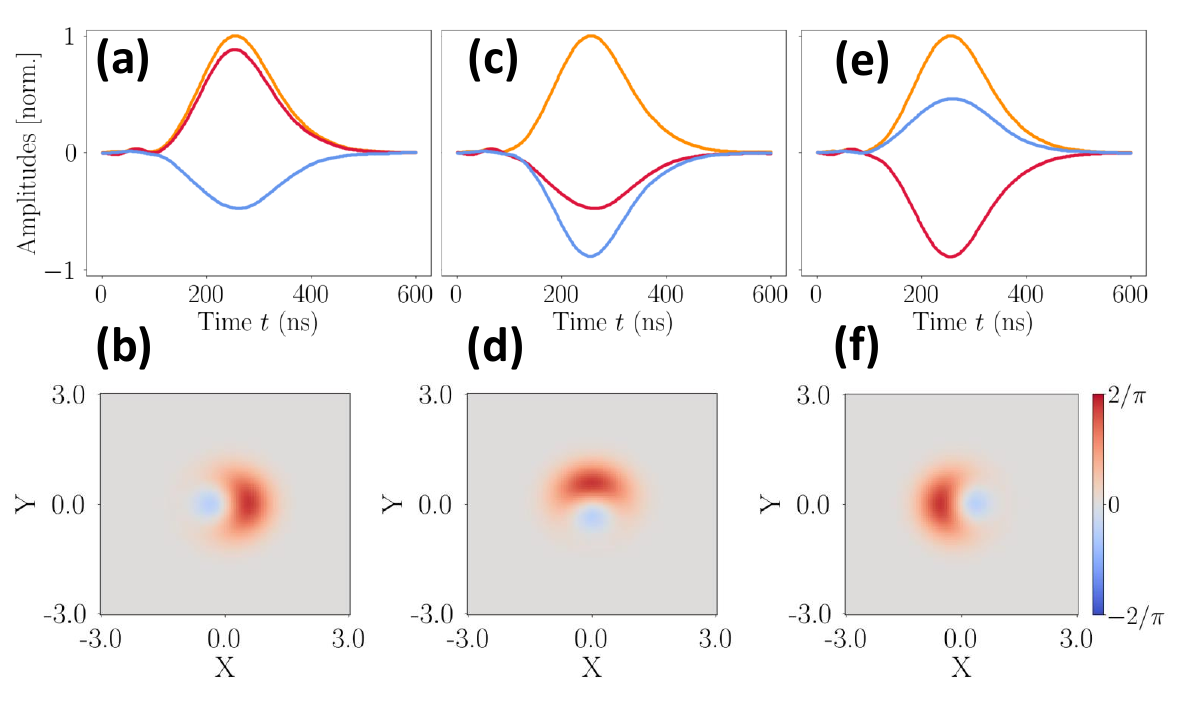}
        \caption{\textbf{The photon phase dependence of the first-order side-band parametric modulation pulse.} (a),(c),(e) are the time-dependent quadratures results(Orange line is $|\braket{\hat{a}_{\textcolor{red}{\text{out}}}(t)}|$, red line is $Re(\braket{\hat{a}_{\textcolor{red}{\text{out}}}(t)})$, blue line is $Im(\braket{\hat{a}_{\textcolor{red}{\text{out}}}(t)})$), and (b),(d),(f) are the Wigner function results. \textbf{(a-b)}The time-dependent quadratures and Wigner function of parametric modulation phase $\theta_{m}=0$. \textbf{(c-d)}The time-dependent quadratures and Wigner function of parametric modulation phase $\theta_{m}=\pi/2$. \textbf{(e-f)}The time-dependent quadratures and Wigner function of parametric modulation phase $\theta_{m}=\pi$.}
        \label{fig:a4}
    \end{figure}
    We test the photon phase dependence of the first-order side-band parametric modulation by prepare the $(\ket{g}+\ket{e})/\sqrt{2}$ state qubit and then emit the photon with different initial phase of the parametric modulation pulse ($\theta_{m}=0,\pi/2,\pi$), then measure the time-dependent quadratures and the Wigner function of the emitted photon, as shown in Fig.~\ref{fig:a4}. The absolute value of quadrature $|\braket{\hat{a}_{\textcolor{red}{\text{out}}}(t)}|$ remains unchanged, while the real and imaginary value $Re(\braket{\hat{a}_{\textcolor{red}{\text{out}}}(t)})$ and $Im(\braket{\hat{a}_{\textcolor{red}{\text{out}}}(t)})$ exchanges absolute values and changes signs in~\ref{fig:a4}(a),(b), change signs in~\ref{fig:a4}(a),(c) representing a photon phase rotation with the phase of parametric modulation. In single-photon subspace, the Wigner function is sensitive to the relative phase of $\ket{0}$ and $\ket{1}$ state which induces a counter-clockwise rotation of Wigner function. The results in Fig.~\ref{fig:a4}(b),(d),(f) further show that the photon phase changes right along with the phase of parametric modulation. The photon phase dependence of the first-order side-band parametric modulation verifies that the effective coupling is a complex coupling $g_{\textcolor{red}{\text{eff}}}(t)=g_{qE}A^{1}_{m}(t)/(2\omega_{m})e^{i\theta_{m}(t)}$, and we can use the phase of parametric modulation $\theta_{m}(t)$ to compensate the phase accumulation of emitted photon.

\section{\MakeUppercase{thermal excitation tests}}
\label{sec:app7}
    \begin{figure}[htbp]
        \centering
        \includegraphics[width=0.45\textwidth]{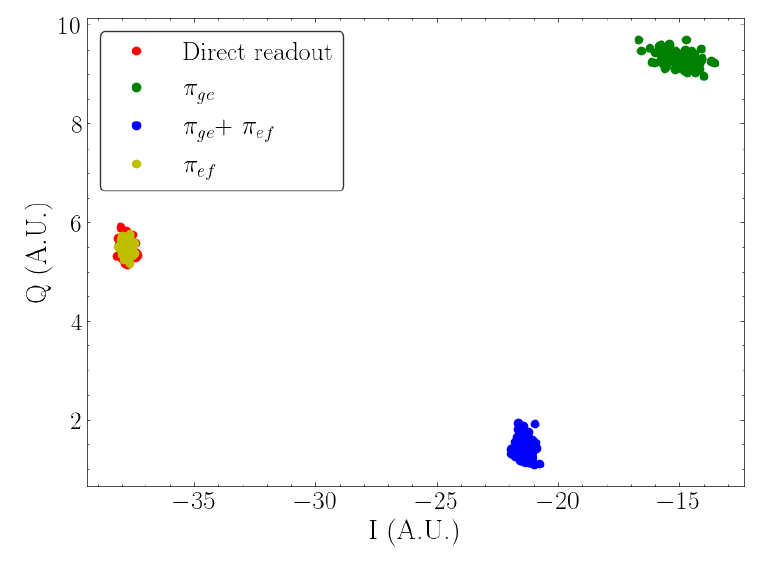}
        \caption{\textbf{The experimental test of thermal fluctuations using high level $\ket{f}$ of qubit and four different control sequences.} Each dot we average for $10^4$ times, there are 100 dots for each control sequence.}
        \label{fig:a5}
    \end{figure}
    We use the method from Ref.~\cite{pechal2016microwave}, which assumes the initial populations of states higher than $\ket{e}$(e.g.$\ket{f}$) are zero. Assume that the eigen-state readout voltages are $v_{g},v_{e},v_{f}$ for $\ket{g},\ket{e},\ket{f}$ states in the absence of thermal excitation, $P_g$, $P_e$ is the initial population of $\ket{g}$ and $\ket{e}$ state. We use four different control pulse sequence to excite the qubit and then readout. The first sequence does not use any control pulse, the second sequence uses the $\pi$ pulse of $\ket{g}$ - $\ket{e}$ state, the third sequence uses the $\pi$ pulse of $\ket{g}$ - $\ket{e}$ state and $\ket{e}$ - $\ket{f}$ state, the fourth sequence uses the $\pi$ pulse of $\ket{e}$ - $\ket{f}$ state. The expected values of readout voltages of the four cases are 
    \begin{align}
    v_1 &= P_g v_g + P_e v_e, \\
    v_2 &= P_e v_g + P_g v_e, \\
    v_3 &= P_e v_g + P_g v_f, \\
    v_4 &= P_g v_g + P_f v_f,
    \end{align}
    where $v_1,v_2$ are the linear combinations of $v_g,v_e$, $v_3,v_4$ are the linear combinations of $v_g,v_f$. Then $v_g$ can be expressed as the linear combinations of both $v_1,v_2$ and $v_3,v_4$, we assume the linear coefficient as $\eta$ and $\lambda$, then we get
    \begin{align}
    v_g = \eta v_2 + (1 - \eta)v_1 = \lambda v_3 + (1 - \lambda)v_4.
    \end{align}
    By solving this equation, we can get 
    \begin{align}
    \eta = \frac{Im \left( (v_3 - v_4)^* v_1 + v_4^* v_3 \right)}{Im \left( (v_2 - v_1)^* (v_3 - v_4) \right)},\\
    \lambda = - \frac{Im \left( (v_2 - v_1)^* v_4 + v_1^* v_2 \right)}{Im \left( (v_2 - v_1)^* (v_3 - v_4) \right)}.
    \end{align}
    And then the thermal population $P_e$ is 
    \begin{align}
    P_e = \frac{\eta}{2\eta - 1} = \frac{\lambda}{2\lambda - 1}.
    \end{align}
    We perform $10^6$ times readout for each control sequence, from the experimental data we get the thermal population $P_e$ is 0.23\% and 0.06\% by $\eta$ and $\lambda$, respectively. Thus, we confirm that the upper-bond of thermal population of our qubit is around 0.2\% corresponding to an effective temperature of 50 mK. The results demonstrate that our device has negligible thermal excitation so that it is not necessary to initialize our qubit before the photon emission process. 

\newpage
\bibliography{OnDemand_Shaped_Photon_By_LFM}

\end{document}